\documentclass[aps,prb,twocolumn]{revtex4}

\usepackage{graphicx}
\usepackage{placeins}
\usepackage{blindtext}
\usepackage{enumitem}
\usepackage{amsfonts}

\usepackage{amsmath,amsthm}

\usepackage[]{bigints}

\begin{document}
\title{Surfaces away from horizons are not thermodynamic}

	\author{Zhi-Wei Wang,${}^{1,2,\dagger}$
and Samuel L.\ Braunstein${}^{2,\ast}$}
	\affiliation{${}^1$College of Physics, Jilin University,
Changchun, 130012, People's Republic of China}
	\affiliation{${}^2$Computer Science, University of York,
York YO10 5GH, United Kingdom}
	\affiliation{${}^\dagger$zhiweiwang.phy@gmail.com}
	\affiliation{${}^\ast$sam.braunstein@york.ac.uk}

\begin{abstract}

Since the 1970's it has been known that black hole (and other) horizons
are truly thermodynamic in nature. More generally, surfaces which are
not horizons have also been conjectured to behave thermodynamically.
Initially, for surfaces microscopically expanded from a horizon to
so-called stretched horizons, and more recently, for more general
ordinary surfaces in the emergent gravity program. To test these
conjectures we ask whether such surfaces satisfy an analogue to the
first law of thermodynamics (as do horizons). For static
asymptotically-flat spacetimes we find that such a first law holds on
horizons. We rigorously prove that this law remains an excellent
approximation for stretched horizons, but perhaps counter-intuitively
this result illustrates the insufficiency of the laws of black hole
mechanics alone from implying truly thermodynamic behavior. For surfaces
away from horizons in the emergent gravity program the first law fails
(with the exception of fully spherically symmetric scenarios) thus
undermining the key thermodynamic assumption of this program.

\end{abstract}

\maketitle

In 1973, Bardeen, Carter, and Hawking derived the laws of black hole
mechanics which are in direct analogy with the laws of thermodynamics
\cite{Bardeen1973}. Together with the discovery of Hawking radiation
\cite{Hawking1975}, the truly thermodynamic behavior of black hole
horizons became well established. Indeed such thermodynamic behavior is
now well accepted for all spacetime horizons, including those due to
accelerated observers \cite{unruh1976,jacobson1995} and cosmological
horizons \cite{gibbons1977}.

Later, other surfaces were also attributed with thermodynamic
properties. Firstly, stretched horizons were claimed to be
thermodynamic, effectively acting as radiating black bodies
\cite{Thorne1986} with a temperature $T=\kappa/(2\pi)$ determined by
their local surface gravity $\kappa$ and an entropy (a `state variable')
associated with a statistical mechanical interpretation of black hole
entropy \cite{Thorne1986, Thorne1985}. An explicit re-derivation of the
laws of black hole mechanics has not been previously carried out for
stretched horizons. More recently, a class of ordinary surfaces has
been conjectured to behave thermodynamically, forming the key assumption
in the emergent gravity program \cite{Verlinde2011}. This thermodynamic
attribution was justified in part by using it in a heuristic derivation
of the full Einstein field equations in static asymptotically-flat
spacetime \cite{Verlinde2011}.

Here we ask whether canonical General Relativity is consistent with the
assumption that such ordinary surfaces can be rigorously seen to behave
thermodynamically. We attack this question by focusing on the analogue
to the first law of thermodynamics. Originally this law was derived in
an analysis that was specialized to the behavior of horizons
\cite{Bardeen1973}. Here we remove this specialization to reveal the
behavior of ordinary surfaces in an analysis of the first law.

\section*{Results}

\begin{figure}[ht]
\centering
\includegraphics[width=0.3\textwidth]{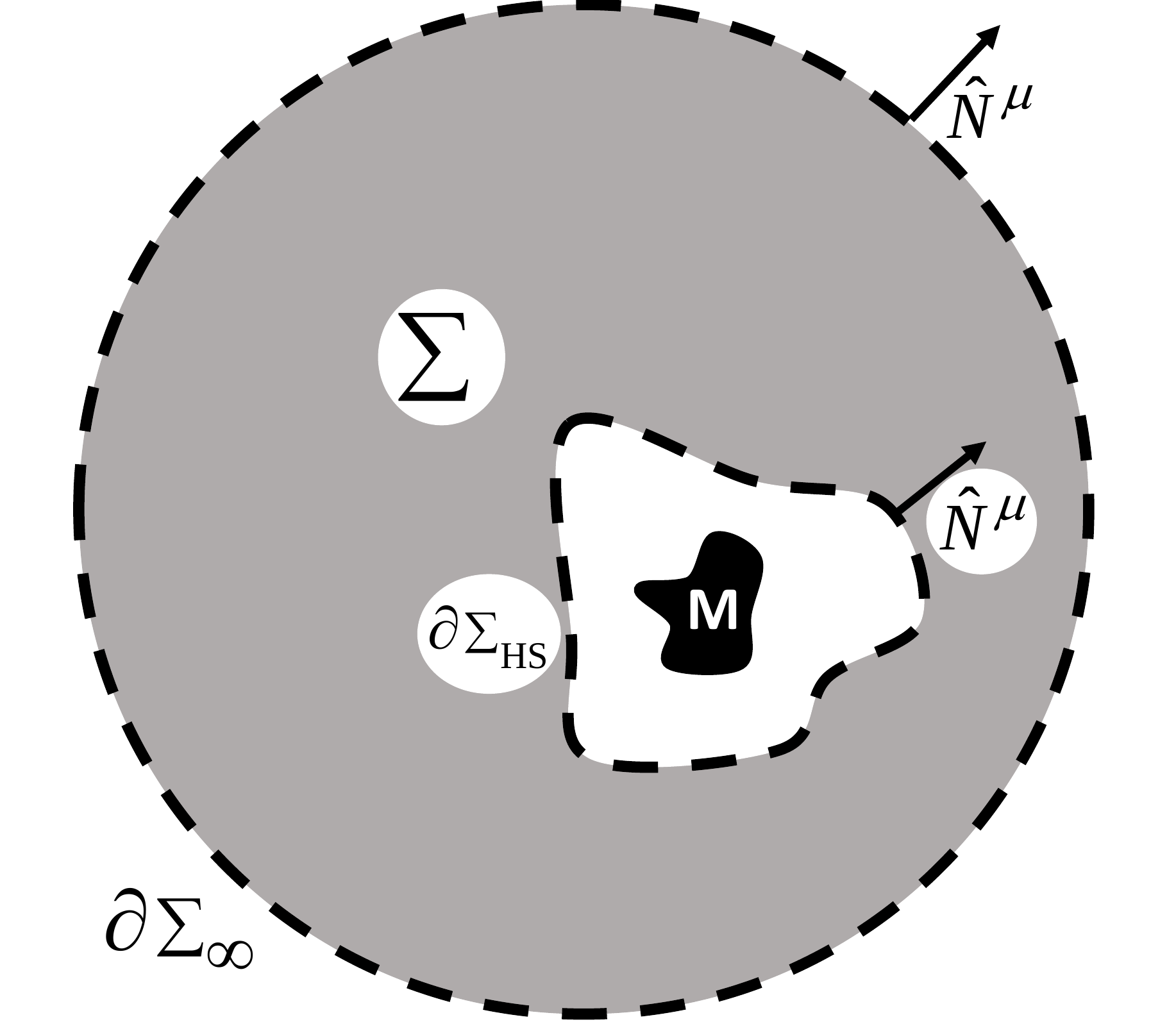}
\caption{Schematic of the spacelike three-dimensional hypersurface of
interest, $\Sigma$, with an inner boundary $\partial \Sigma_{\text{in}}$
and a boundary at infinity $\partial \Sigma _\infty$. Here $\hat N^\mu$
is the spacelike 4-vector normal to the boundaries of $\Sigma$ (note the
direction convention on the inner boundary). We assume a general mass
distribution within the inner boundary and no matter outside it.}
\label{fig1}
\end{figure}

For a static asymptotically-flat spacetime with timelike Killing vector
$K^\mu$ one may derive the {\it total\/} gravitating mass $M$ as an
integral over a spacelike hypersurface $\Sigma$ that is truncated (or
bounded) internally by an ordinary 2-surface
$\partial\Sigma_{\text{in}}$ (see Fig.~\ref{fig1}) 
\begin{equation}
    M =  \frac{1}{4 \pi} \int _\Sigma \!R_{\mu \nu}  K^\mu \,
    \hat T^\nu \, \sqrt{|\gamma^{(\Sigma)}| } \; d^3x
+ \frac{1}{4 \pi}
    \int _{\partial\Sigma_{\text{in}} }\!\!\!\!\kappa\; dA .
    \label{massss}
\end{equation}
(See the Supplementary Information for a detailed derivation and
definition of each term.) This expression is a straightforward extension
of that used in 1973 by Bardeen et al.\ \cite{Bardeen1973} in their
derivation of the first law of thermodynamics for black holes, though
there the internal boundary was a horizon. Here $\kappa$ is a natural
extension of the surface gravity for non-rotating spacetimes.

Following the results for horizons \cite{Bardeen1973}, it is {\it
tempting\/} to seek to interpret $\kappa/(2\pi)$ from Eq.~(\ref{massss})
as the local temperature at any point along an arbitrary 2-surface
$\partial \Sigma_{\text{in}}$. However, this would be unsatisfactory if
true for arbitrary surfaces, since this local temperature would {\it
not\/} be in thermal equilibrium with an actual physical screen held
fixed at the same location; the temperature now coming from the Unruh
effect \cite{unruh1976} and the local proper
acceleration required to keep each portion of the screen stationary.
Only for surfaces of constant Newtonian gravitational potential $\phi$,
where the proper acceleration of a stationary observer and the local
normal to the surface are parallel, is such thermal equilibrium possible
(see Supplementary Information). Thus the temptation of such a
thermodynamic interpretation should be restricted to the family of
ordinary surfaces satisfying $\phi={\text{constant}}$.

Indeed, this restricted temptation appears to have been satisfied in the
emergent gravity program, where for static asymptotically-flat
spacetimes, ordinary surfaces of constant $\phi$ are dubbed holographic
screens and are claimed to have a local temperature \cite{Verlinde2011}
given by $T=\kappa/(2\pi)$ and even to possess a `state variable'
quantifying the number of `bits' on the screen. These putative
thermodynamic properties are then used in a heuristic derivation of the
full Einstein field equations \cite{Verlinde2011}. {\it If correct, such
a claim would mean that the emergent gravity program would already
subsume many decades of results associated with full General Relativity
in this setting}.

Here we test this thermodynamic assumption by asking whether
perturbations of Eq.~(\ref{massss}) reproduce the first law of
thermodynamics. After all, thermodynamics is primarily a theory about
how energy transforms under change, and this aspect of the theory is
embodied in the first law. In the simplest case, where the hypersurface
$\Sigma$ is empty of matter, this law should read
\begin{equation}
\delta M = \frac{1}{8 \pi} \int _{\partial\Sigma_{\text{in}} }
\!\!\!\!\kappa \; \delta (dA).
\label{1st}
\end{equation}

We start by following Bardeen et al.'s original analysis
\cite{Bardeen1973}, generalizing it where necessary to deal with a
boundary $\partial\Sigma_{\text{in}}$ which is an {\it arbitrary\/}
ordinary surface instead of a horizon. Under diffeomorphic metric
perturbations we find (see Supplementary Information)
\begin{eqnarray}
 \delta M \!&=&\!
\frac{-1}{32 \pi} \int _{\partial\Sigma_\text{in} }
\!\!\!\bigl[  4 \delta \theta ^{(l)} +  ( k_3+k_6) \theta ^{(l)}
+  ( k_3-k_6) \sigma_+ ^{(l)} \nonumber \\
&&~~~~+  (k_4+k_5)
\sigma_\times ^{(l)} \bigr]\, {\cal N} dA
+ \frac{1}{8 \pi} \int _{\partial\Sigma_\text{in} }
\!\!\!\!\kappa \, \delta (dA).
\label{firstlaw2}
\end{eqnarray}
Here $\theta ^{(l)}$ and $\sigma_j^{(l)}$ ($j=+,\times$) are the
expansion and shears of null normal congruences of geodesics, 
the change in the expansion under the diffeomorphism is given by
\begin{equation}
\delta \theta ^{(l)}=
-\frac{k_2}{2} \theta ^{(l)} + \frac{1}{2} (k_3 + k_6)_{;\rho } \hat N^\rho,
\end{equation}
and the $k_j$ are functions corresponding to independent components of
the metric perturbation. As the expansion and shears vanish identically
on the horizon \cite{hawking1973}, we see that Eq.~(\ref{firstlaw2})
trivially reduces to the first law, Eq.~(\ref{1st}), thus reproducing
the famous 1973 result \cite{Bardeen1973}. Similarly, it follows
straightforwardly that for surfaces sufficiently close to the horizon
(so-called stretched horizons), the corrections to the first law will be
negligible.

So far we have assumed that the inner boundaries before and after the
diffeomorphic perturbation are arbitrary. But could the perturbed
boundary be chosen in a {\it specific\/} manner so as to cause the unwanted
terms in Eq.~(\ref{firstlaw2}) to vanish? As already noted, holographic
screens correspond to surfaces of constant Newtonian potential
$\phi={\text{constant}}$. Thus, the perturbed screen relies on a
specification of the constant $\delta \phi$. It is easy to show that
$\delta\phi=\frac{1}{2}k_1$ (see Supplementary Information), where $k_1$
is a metric perturbation of which the unwanted terms in
Eq.~(\ref{firstlaw2}) are wholly independent. Thus, the ordinary
surfaces used within the emergent gravity program cannot generally
satisfy the first law, Eq.~(\ref{1st}).

One caveat to this claim comes when we consider a fully spherically
symmetric scenario; where both the initial spacetime and screen are
spherically symmetric, so the initial shears $\sigma_j^{(l)}$ vanish,
and also the final spacetime and screen are spherically symmetric,
placing further constraints on the $k_j$. In this case, Birkhoff's
theorem \cite{Birkhoff1923} for spherically symmetric metrics imposes
extra constraints between the metric components so that a perturbed
screen may always be chosen so as to satisfy the {\it form\/} of the
first law \cite{Chen11}. However, as noted above, this form will not be
preserved under arbitrary metric perturbations.

\section*{Discussion}




The implications of our results are now described for (i) stretched
horizons, and (ii) ordinary surfaces.

(i) Stretched horizons have long been considered to act as black bodies
\cite{Thorne1986}, effectively radiating with a temperature $\kappa /(2
\pi)$. Thus, our demonstration that they also satisfy the first law to
an excellent approximation hardly seems surprising. Nevertheless, we do
not believe that our result here should be interpreted as implying that
the surfaces corresponding to stretched horizons themselves should be
imbued with actual thermodynamic properties.

In particular, we may consider an alternative spacetime, identical from
the stretched horizon outward, but instead of a horizon, we consider an
infinitesimal shell of matter just outside what would correspond to its
Schwarzschild radius were the shell to collapse further, yet still
within the `stretched horizon'. In this latter spacetime, there is no
horizon and hence no Hawking radiation. Notwithstanding this, our work
proves that the `stretched horizon' still closely satisfies the first
law.

We conclude from this that the laws of black hole mechanics are {\it
not\/} sufficient in themselves to guarantee whether any particular
surface is truly thermodynamic in nature. For stretched horizons, we
interpret this reasoning to imply that their full thermodynamic behavior
is only inherited from the presence of an underlying horizon, but is not
intrinsic to stretched horizons themselves. This conclusion appears to
mimic the initial reluctance of general relativists\cite{Bardeen1973}
from accepting black hole horizons as truly thermodynamic despite the
deep analogy to thermodynamics uncovered in the laws of black hole
mechanics. By contrast, these laws should still be considered a
necessary condition.

(ii) Our analysis further rigorously shows that the family of ordinary
surfaces called holographic screens will generally {\it not\/} obey a
first law of thermodynamics, in contrast to the long-standing result for
horizons \cite{Bardeen1973}. (Other families would not even be in
thermal equilibrium with a physical surface at the same location.)
Recall that the first law is more general than thermodynamics: the
`temperature' is merely an integrating factor relating changes in energy
to changes in some state variable (entropy in the case of
thermodynamics). Failure of the first law means that the putative state
variable is not a variable of state at all. Therefore, even in static
asymptotically-flat spacetimes, where the emergent gravity program
claims to derive the full Einstein field equations, our results show
that the key assumption of this program is actually inconsistent with
General Relativity.

\section*{Methods}

In order to attempt to derive a first law for ordinary surfaces we
closely follow in the footsteps of Bardeen, Carter and Hawking's 1973
classic paper \cite{Bardeen1973}. The first step is to obtain an
integral equation for the net energy in a static system,
Eq.~(\ref{massss}), where instead of an inner boundary located at a
black hole horizon, this boundary is an ordinary surface. Next, we
consider small `changes' in the net energy corresponding to shifting to
a parametrically nearly solution to the Einstein field equations. This
`differential' version is determined by studying the behavior of the net
energy under spacetime diffeomorphisms of the initial metric
\cite{Bardeen1973}. As in Bardeen et al., ``gauge'' freedom in the
choice of coordinates is used to ensure that the hypersurfaces before
and after the diffeomorphism are covered by identical sets of
coordinates.

Our analysis is limited to static asymptotically-flat solutions, with
zero shift vector, $\beta^\mu=0$. For simplicity, we assume that the
spacetime of interest is non-rotating, and that there is no matter {\it
exterior\/} to the holographic screen ($T^{\mu\nu}=0$). We work
throughout in natural units where $G=c=\hbar=k_B=1$. Full and extensive
details of the analysis are provided in the Supplementary Information.

\section*{Appendix}

\subsection{List of key symbols}
We set $G=c=\hbar=k_B=1$
throughout. A list of key symbols used in the Appendix:
\begin{description}[style=multiline,labelindent=10pt,%
itemsep=0mm,align=left,leftmargin=1.5cm]
    \item [$M$] mass
    \item [${\cal N}$] lapse function ($=1/\hat T^t$)
    \item [$\beta^\mu$]  shift vector ($=0$ throughout)
    \item [$g_{\mu\nu}$]  metric
    \item [$T^{\mu \nu}$] energy-momentum tensor ($=0$ from Eq.(\ref{Ricci}))
    \item [$R_{\mu \nu}$]  curvature tensor
    \item [$R$] Ricci scalar
    \item [$\Sigma$] 3-dimensional hypersurface in this paper
    \item [$\Sigma_\text{EG}$] 3-dimensional hypersurface in Ref.~\onlinecite{Verlinde2011}
    \item [$\partial \Sigma$] boundary of $\Sigma$
    \item [$\partial \Sigma_{\infty}$]  outer boundary of $\Sigma$
    \item [$\partial \Sigma_\text{in}$]  inner boundary of
$\Sigma$  (ordinary surface)
    \item [$\partial \Sigma_\text{BH}$]  inner boundary
of $\Sigma$ (black hole horizon)
    \item [$\partial \Sigma_\text{HS}$]  inner boundary
of $\Sigma$ (holographic screen)
    \item [$\gamma^{(\Sigma)}$] induced metric on $\Sigma$
    \item [$\gamma^{(\partial\Sigma)}$] induced metric on $\partial\Sigma$
    \item [$dA$] area element ($=\sqrt{\gamma^{(\partial\Sigma)}}\,dy^2$)
    \item [$\hat T^\mu$] timelike unit normal vector of $\Sigma$
    \item [$\hat N^\mu$] spacelike unit normal vector
of $\partial \Sigma$
    \item [$\kappa$]  generalized surface gravity, Eq.~(\ref{def})
    \item [$K^\mu$] Killing vector of the static spacetime ($=\partial_t$)
    \item [$h_{\mu \nu}$] diffeomorphic variation of the 
metric ($=\delta g_{\mu \nu}$)
    \item [$\theta^{(l)}$] expansion of the outgoing null normal vector $l^\mu$
    \item [$\sigma_+ ^{(l)} ,\, \sigma_\times^{(l)}$] shears of the
outgoing null normal vector $l^\mu$
    \item [$\sigma_P ^{(l)}$] `principle shear' of the outgoing null normal
vector $l^\mu$
\end{description}

\subsection{Integral expression for net energy}

Consider a static spacetime with a Killing vector 
$K^\mu = \partial_ t= (1,0,0,0)$, with $K^\mu K_\mu=-1$ at spatial 
infinity. The Killing equation implies that
\begin{equation}
    K_{\mu ; \nu}= K_{[\mu ; \nu]}\equiv\frac{1}{2}
(K_{\mu ; \nu}- K_{\nu ; \mu}) \; .
    \label{Killing 1}
\end{equation}
Now recall that permuting the order of a pair of covariant derivatives
acting on a 4-vector $A^\mu$ may be expressed in terms of the
Riemann curvature tensor as \cite{carroll2004}
\begin{equation*}
    {A^\mu}_{;\alpha \beta} - {A^\mu}_{;\beta \alpha } 
    = - {R^\mu}_{ \nu \alpha \beta} A^\nu \; .
\end{equation*}
Contracting the indices $\mu$ and $\alpha$ reduces this to
an expression in terms of the Ricci tensor
\begin{equation}
    {A^\mu}_{;\mu \beta} - {A^\mu}_{;\beta \mu } 
    = - R_{ \nu \beta} A^\nu \; .
    \label{curve}
\end{equation}

Since the Killing vector is anti-symmetric we must have
${K^\mu}_{;\mu}=0$ and we immediately find that
\begin{equation}
    {K^\mu}_{;\beta \mu } = R_{ \nu \beta} K^\nu.
    \label{Killing 2}
\end{equation}
Integrating this over a spacelike hypersurface $\Sigma$, yields
\begin{equation}
   \int _\Sigma  {K^\mu}_{; \beta \mu } \, \hat T^\beta \, 
   \sqrt{|\gamma^{(\Sigma)} | } \, d^3x= 
\int _\Sigma R_{ \nu \beta} K^\nu \, 
   \hat T^\beta \, \sqrt{|\gamma^{(\Sigma)} | } \, d^3x
    \label{o}
\end{equation}
here $\hat T^\mu$ is the timelike unit 4-vector normal to $\Sigma$, so 
$\hat T^\mu \hat T_\mu=-1$.

The hypersurface is assumed to have an outer boundary at spatial infinity
$\partial \Sigma_\infty$, and an inner boundary $\Sigma_\text{in}$
(see Fig.~1). In the
original work of Bardeen et al.\ \cite{Bardeen1973}, this inner boundary
corresponded to the black hole's horizon $\partial \Sigma_{\text{BH}}$.
Here we generalize this by taking it to be an arbitrary closed 2-surface 
$\partial \Sigma_{\text{in}}$. The boundary of the
hypersurface is assumed to be oriented, with unit normal $\hat N^\mu$ (see
Fig.~1), so $\hat N_\mu \hat N^\mu=1$ and $\hat N^\mu \hat T_\mu=0$.

Recalling Stokes's theorem
for an anti-symmetric tensor $F^{\mu\nu}$ \cite{carroll2004}
\begin{equation}
    \int_\Sigma \hat T_\mu {F^{\mu \nu}}_{;\nu}
\sqrt{|\gamma^ {(\Sigma)} |} dx^{n-1} = \int_{\partial \Sigma}
\hat T_\mu F^{\mu \nu} \hat N_\nu \sqrt{|\gamma^{(\partial \Sigma)} |}
dy^{n-2},
    \label{stokes}
\end{equation}
and applying it to the left-hand-side of 
Eq.~(\ref{o}) we find
\begin{widetext}
\begin{eqnarray}
   \int _ {\partial \Sigma _\infty } {K^\mu}_{; \beta } \,  \hat T^\beta 
   \hat N_\mu 
    \sqrt{|\gamma^{(\partial \Sigma _\infty)} |  }  \;d^2y -\int _ {\partial \Sigma_{\text{in}}} {K^\mu}_{; \beta } \,  \hat T^\beta 
   \hat N_\mu \, 
    \sqrt{|\gamma^{(\partial \Sigma_{\text{HS}})} |  }\;  d^2y
   =  \int _\Sigma R_{ \nu \beta} K^\nu \, \hat T^\beta \,
\sqrt{|\gamma^{(\Sigma)} | } \, d^3x  .
    \label{o1}
\end{eqnarray}
\end{widetext}

At this stage, we wish to generalize the concept of surface gravity
as a quantity defined anywhere. Assuming that the surface
$\partial \Sigma$ is non-rotating (corresponding to zero angular velocity
of the spacetime itself) , we may interpret the integrand of
the integral on the boundary in Eq.~(\ref{o1}) to be the surface
gravity, so
\begin{equation}
\boxed{
\kappa \equiv {K^\mu}_{;\nu} \hat T^\nu \hat N_\mu .
}
\label{def}
\end{equation}
It is worth noting that ${\kappa}/({2\pi})$ is precisely the
formula Verlinde gives (his Eq.~(5.3) of Ref.~[\onlinecite{Verlinde2011}]) for what he calls
the local temperature of the holographic screen (ordinary surfaces of constant
Newtonian potential $\phi$) as measured with respect
to a reference point at spatial infinity.

This definition of surface gravity allows us to naturally extend the
original 1973 analysis away from black hole horizons. In particular,
the left-hand-side of Eq.~(\ref{o1}) reduces to
\begin{eqnarray}
\int _ {\partial \Sigma _\infty }  \kappa \,
    \sqrt{|\gamma^{(\partial \Sigma _\infty)} |  }  \;d^2y - \int _ {\partial \Sigma_{\text{in}}} \kappa\,
    \sqrt{|\gamma^{(\partial \Sigma_{\text{HS}})} |  }\;  d^2y .
\end{eqnarray}
The integral over $\partial \Sigma_\infty$ reduces to the Komar expression
for the {\it total\/} gravitating mass within the system, $M$,
\cite{carroll2004} leading to
\begin{eqnarray}
    \boxed{
    M =  \frac{1}{4 \pi} \int _\Sigma R_{\mu \nu}  K^\mu \,
    \hat T^\nu \, \sqrt{|\gamma^{(\Sigma)}| } \; d^3x
+ \frac{1}{4 \pi}
    \int _{\partial\Sigma_{\text{in}} }\!\!\kappa\; dA \;, } \nonumber \\
    \label{massssAp}
\end{eqnarray}

Were we to consider spherically symmetric case, Eq.~(\ref{massssAp}) would reduce to
\begin{equation}
    M =  \frac{1}{4 \pi} \int _\Sigma R_{\mu \nu}  K^\mu \, 
    \hat T^\nu \, \sqrt{|\gamma^{(\Sigma)}| } \, d^3x
+ \frac{\kappa}{4 \pi} A.
\label{netMass}
\end{equation}
(This is exactly the first law given by Bardeen et al.\ \cite{Bardeen1973}
taking angular velocity of $\partial \Sigma$ to vanish).

Just to emphasize what this represents, here the hypersurface, $\Sigma$,
extends from an arbitrary inner boundary, $\partial \Sigma_{\text{in}}$,
out to spatial infinity. Thus, the generalized surface gravity,
$\kappa$, and the area, $A$, are those associated with the inner
boundary itself (rather than any horizon).

Eq.~(\ref{massssAp}) has exactly the same form as the conventional formula
for the total mass of the system \cite{Bardeen1973} but extended to an
arbitrary 2-dimensional surface (instead of a horizon). Finally, note
that the matter inside the inner boundary need {\it not\/} be associated
with a black hole, it may be ordinary matter, with no horizon present at
all. Thus, were inner boundaries found to have thermodynamic properties
(i.e., a well-defined entropy and temperature), it would {\it not\/} be
because such properties were inherited from a real horizon behind the
screen.

\subsection{Differential ``first law''}

The above straightforward generalization, especially in the spherically
symmetric case, for net energy on a hypersurface might appear to
suggest that a temperature and entropy can actually be defined for any
surface by
\begin{equation}
T=\frac{\kappa}{2\pi},\qquad \qquad S=\frac{A}{4}.
\end{equation}
However, such quantities need to behave
thermodynamically. In particular, for our static system, the net energy
$E$, should admit changes which behave as
\begin{equation}
\delta E= T\delta S,
\end{equation}
(ignoring work terms) so that the temperature would be acting as an
integrating factor relating changes in the (state function) entropy to
changes in the energy. In other words, we must show that such changes
lead to the expected form of the first-law of thermodynamics. Again here we
follow in the footsteps of the original analysis and consider changes
corresponding to parametric differences between diffeomorphicly nearby
solutions. In particular, we will consider two nearby configurations
corresponding to the metrics
\begin{equation}
g_{\mu\nu},\qquad \qquad g'_{\mu\nu}=g_{\mu\nu}+h_{\mu\nu},
\end{equation}
where $h_{\mu\nu}\equiv \delta g_{\mu\nu}=
- g_{\mu\sigma} g_{\nu \tau} \delta g^{\sigma\tau}$, i.e.,
$\delta g^{\sigma\tau} = -h^{\sigma\tau}$.

As with the original analysis and without loss of generality, we may
assume that for the two diffeomorphicly related configurations, the 
hypersurfaces $\Sigma$ and $\Sigma '$ are described by identical sets 
of coordinates; this is always possible due 
to ``gauge'' freedom in the choice of coordinate systems \cite{Bardeen1973}. 
Henceforth we label both by $\Sigma$. Similarly, for their boundaries $\partial \Sigma$. Further, as in \cite{Bardeen1973} we likewise 
assume that both configurations have the same Killing vector, so 
\begin{equation}
\delta K^\mu = 0, \qquad\qquad \delta K_\mu = h_{\mu \nu} K^\nu .
\end{equation}
Finally, it will be sufficient for our purposes to
consider only the case where there is no matter on $\Sigma$ itself,
so $T^{\mu\nu}=0$ there. Geometrically, this corresponds to all the matter
lying behind or within the inner boundary $\partial \Sigma_{\text{in}}$
(see Fig.~1).

\begin{widetext}
In order to consider diffeomorphisms which need not respect spherical
symmetry, we return to Eq.~(\ref{massssAp}).
Using the Einstein field equations we start by rewriting this
integral formula as
\begin{eqnarray}
    M =  \int _\Sigma ( 2\, T_{\mu \nu} +\frac{1}{8 \pi} R\,
 g _{\mu \nu }) K^\mu \, \hat T^\nu \, \sqrt{|\gamma^{(\Sigma)}| } \, d^3x
+ \frac{1}{4 \pi} \int _{\partial\Sigma_\text{in} } \kappa  \; dA  \; ,
\end{eqnarray}
Recall $K^\mu = \partial_t$ and $\hat T^\mu$ is normal to $\Sigma$, so $K^\mu = {\cal N} \hat T^\mu + \beta^\mu$ where $\beta^\mu$ is the shift vector and ${\cal N}=1/\hat T^t$ is
the lapse function \cite{gourgoulhon2012}. Assuming a zero shift vector $\beta^\mu =0$, then $\hat T^\mu= \hat T^t K^\mu$ and $\hat T_\mu= \hat T^t K_\mu$. Since $ {\cal N} \sqrt{|\gamma^{(\Sigma)} |}= \sqrt{-g }$ on the
hypersurface \cite{gourgoulhon2012}, the variation of the Ricci scalar term may be computed as
\begin{eqnarray}
    && \frac{1}{8 \pi} \int _\Sigma  \delta (R \sqrt{|\gamma^{(\Sigma)} |}
\,K^\beta \, \hat T_\beta )  \, d^3x  \nonumber  \\
    &=&\frac{1}{8 \pi} \int _\Sigma  \delta (R \,
{\cal N} \sqrt{|\gamma^{(\Sigma)} |}
\,\hat T^\beta \, \hat T_\beta)  \, d^3x   \nonumber  \\
    &=&\frac{1}{8 \pi} \int _\Sigma  \delta (R \, \sqrt{-g})
 \,\hat T^\beta \, \hat T_\beta  \, d^3x   \nonumber  \\
    &=& -\frac{1}{8 \pi} \int _\Sigma  \Bigl ( (R_{\mu \nu}-\frac{1}{2}
g_{\mu \nu}R ) h^{\mu \nu} - (g^{\mu\nu} \delta {\Gamma ^\alpha}_{\mu \nu }
-g^{\mu\alpha } \delta {\Gamma ^\lambda}  _{\lambda  \mu })_{;\alpha }
 \Bigl ) \, K^\beta \, \hat T_\beta \,
\sqrt{|\gamma^{(\Sigma)} | } \, d^3x \; .
    \label{VariationRicci}
\end{eqnarray}
where in the last step we have used the well-known result
that \cite{bertschinger2002}
\begin{equation}
\delta (R \sqrt{-g}) =-\Bigl( (R_{\mu \nu}-\frac{1}{2}g_{\mu \nu}R )
h^{\mu \nu} - (g^{\mu\nu} \delta {\Gamma ^\alpha} _{\mu \nu }
-g^{\mu\alpha } \delta {\Gamma ^\lambda} _{\lambda  \mu })_{;\alpha }
\Bigr) \sqrt{-g } .
\end{equation}

\noindent
{\bf Lemma 1:} $-(g^{\mu\nu} \delta {\Gamma ^\alpha} _{\mu \nu }
-g^{\mu\alpha } \delta {\Gamma ^\lambda}  _{\lambda  \mu })_{;\alpha }
=2\, {{h^{\mu}}_{[\mu ; \nu]}}^{;\nu}$,
a result quoted in
Ref.~\cite{Bardeen1973}, there without proof.

\noindent
{\bf Proof:}

Since \cite{palatini1919}
\begin{eqnarray}
   \delta \Gamma_{\mu \nu}^\alpha = \frac{1}{2} g^{\alpha \rho } ( h_{\mu \rho ;\nu } + h_{\nu \rho ;\mu }-h_{\mu \nu ;\rho } ) \;,
   \label{de00}
   \end{eqnarray}
we have   
 \begin{eqnarray}
    -(g^{\mu\nu} \delta {\Gamma ^\alpha} _{\mu \nu }-g^{\mu\alpha }
\delta {\Gamma ^\lambda}  _{\lambda  \mu })_{;\alpha } 
    &=& \Bigl( g^{\mu\alpha } \frac{1}{2} g^{\lambda \rho } (  h_{\mu \rho ;\lambda } + h_{\lambda \rho ;\mu }-h_{\mu \lambda ;\rho }  ) -   g^{\mu\nu } \frac{1}{2} g^{\alpha \rho } ( h_{\mu \rho ;\nu } + h_{\nu \rho ;\mu }-h_{\mu \nu ;\rho } ) \Bigr)_{;\alpha } 
    \nonumber  \\
    &=& \frac{1}{2} \Bigl( g^{\mu\alpha }  {h^\rho}_{\rho ;\mu } -   g^{\mu\nu } ({h^\alpha}_{\mu ;\nu } + {h^\alpha}_{\nu ;\mu } - {h_{\mu \nu}}^{;\alpha} ) \Bigr)_{;\alpha } 
    \nonumber  \\
    &=& \frac{1}{2} \Bigl( {h_\rho}^{\rho ;\alpha } - ({h^{\alpha \nu}}_{;\nu } + {h^{\alpha \mu}}_{;\mu } - {h_\mu}^{\mu ;\alpha} ) \Bigr)_{;\alpha }
    \nonumber  \\
    &=& \frac{1}{2} ( 2 \,{h_\rho}^{\rho ;\alpha } - 2\, {h^{\alpha \mu}}_{;\mu } )_{;\alpha }
    \nonumber  \\
    &=& 2\, {{h^{\mu}}_{[\mu ; \nu]}}^{;\nu} .
\end{eqnarray}
This completes the proof of Lemma 1.
\qed

\vskip 0.4in

Using Lemma 1, the variation of the Ricci scalar term becomes
\begin{eqnarray}
    -\frac{1}{8 \pi} \int _\Sigma  \Bigl ( (R_{\mu \nu}-\frac{1}{2}
g_{\mu \nu}R ) h^{\mu \nu} + 2 {{h^{\mu}}_{[\mu ; \nu]} }^{;\nu}
 \Bigl ) \, K^\beta \, \hat T_\beta \,
\sqrt{|\gamma^{(\Sigma)}|} \, d^3x
=-\frac{1}{4 \pi} \int _\Sigma  
 {{h^{\mu}}_{[\mu ; \nu]} }^{;\nu} K^\beta \, \hat T_\beta \,
\sqrt{|\gamma^{(\Sigma)}|} \, d^3x
    \label{Ricci}
\end{eqnarray}
since the first term is zero if we assume $T^{\mu \nu}=0$ on $\Sigma$
outside the holographic screen. 

\vskip 0.3in

\noindent
{\bf Lemma 2:} $ {{h^{\mu}}_{[\mu ; \nu]} }^{;\nu} K^\beta
=  ( K^\beta {h_\mu}^{[\mu;\nu]}
- K^\nu {h_\mu}^{[\mu;\beta]} )_{;\nu}$, a result quoted in
Ref.~\cite{Bardeen1973}, there without proof.

\vskip 0.1in

\noindent
{\bf Proof:} Expanding out the right-hand-side (rhs) of the claim
in Lemma 2, we get
\begin{eqnarray}
 \text{rhs} = {{h_\mu}^{[\mu;\nu]} }_{;\nu}  K^\beta + {h_\mu}^{[\mu;\nu]}
 { K^\beta }_{;\nu} -  {{h_\mu}^{[\mu;\beta]}}_{;\nu} K^\nu \;\; ,
 \label{le2}
\end{eqnarray}
since ${{h_\mu}^{[\mu;\nu]} }_{;\nu}  K^\beta = {{h^{\mu}}_{[\mu ; \nu]}
}^{;\nu} K^\beta$, Eq.~(\ref{le2}) reduces to
\begin{eqnarray}
{{h^{\mu}}_{[\mu ; \nu]} }^{;\nu} K^\beta + {h_\mu}^{[\mu;\nu]}
{ K^\beta }_{;\nu} -
 {{h_\mu}^{[\mu;\beta]}}_{;\nu} K^\nu 
= {{h^{\mu}}_{[\mu ; \nu]} }^{;\nu} K^\beta -\pounds_K
({h_\mu}^{[\mu;\beta]}) = {{h^{\mu}}_{[\mu ; \nu]}
}^{;\nu} K^\beta = \text{lhs} \;.
\end{eqnarray}
The Lie derivative along $K^\mu$ vanishes since the pair of
diffeomorphicly related metrics are assumed static \cite{poisson2004}.
This completes the proof of Lemma 2.
\qed

\vskip 0.5in

Applying Lemma 2 to Eq.~(\ref{Ricci}), the variation of the term
involving the Ricci scalar reduces to
\begin{eqnarray}
    -\frac{1}{4 \pi}\int _\Sigma ( K^\beta {h_\mu}^{[\mu;\nu]}
- K^\nu {h_\mu}^{[\mu;\beta]} )_{;\nu} \, \hat T_\beta \,
\sqrt{|\gamma^{(\Sigma)}| } \, d^3x .
\end{eqnarray}
Thus, the variation in the total mass may be written
\begin{eqnarray}
   \delta M = -\frac{1}{4 \pi}\int _\Sigma ( K^\beta {h_\mu}^{[\mu;\nu]}
- K^\nu {h_\mu}^{[\mu;\beta]} )_{;\nu} \, \hat T_\beta \,
\sqrt{|\gamma^{(\Sigma)}| } \, d^3x
+ \frac{1}{4 \pi} \int _{\partial\Sigma_\text{in} }
\delta \kappa  \; dA + \frac{1}{4 \pi} \int _{\partial\Sigma_\text{in} }
\kappa  \; \delta (dA) .
\label{eq26}
\end{eqnarray}
Since the term inside the bracket is an anti-symmetric tensor, we may use Stokes's theorem, Eq.~(\ref{stokes}), to obtain
\begin{eqnarray}
   \delta M = &&-\frac{1}{4 \pi}\int _{\partial \Sigma_\infty }
( K^\beta {h_\mu}^{[\mu;\nu]} - K^\nu {h_\mu}^{[\mu;\beta]} )
\hat N_\nu \, \hat T_\beta \,
\sqrt{|\gamma^{({\partial \Sigma_\infty })} |} \, d^2y 
\nonumber \\
&&+ \frac{1}{4 \pi}\int _{\partial \Sigma_\text{in} }
( K^\beta {h_\mu}^{[\mu;\nu]} - K^\nu {h_\mu}^{[\mu;\beta]} )
\hat N_\nu \, \hat T_\beta \,
\sqrt{|\gamma^{({\partial \Sigma_\text{in} })} |} \, d^2y
+ \frac{1}{4 \pi} \int _{\partial \Sigma_\text{in} }
\delta \kappa  \; dA
+ \frac{1}{4 \pi} \int _{\partial\Sigma_\text{in} } \kappa  \; \delta (dA),
\label{uuu}
\end{eqnarray}
where the boundary has be
split into the inner boundary $\partial \Sigma_\text{in}$ and the
boundary at infinity $\partial \Sigma_\infty$. The contribution for
the term at infinity may be evaluated using the notation of tensorial
volume elements
\cite{Wald1984} as
\begin{eqnarray}
    - \frac{1}{4 \pi}\int _{\partial \Sigma_\infty }
( K^\beta {h_\mu}^{[\mu;\nu]}
- K^\nu {h_\mu}^{[\mu;\beta]} ) \hat N_\nu \, \hat T_\beta \,
\sqrt{|\gamma^{(\partial \Sigma_\infty)} |} \, d^2y 
&=&  - \frac{1}{4 \pi}\int _{\partial \Sigma_\infty }
K^\beta {h_\mu}^{[\mu;\nu]} (\hat T_\beta \, \hat N_\nu \,
-\hat T_\nu \, \hat N_\beta )
\sqrt{|\gamma^{(\partial \Sigma_\infty)} |} \, d^2y
\nonumber \\
&=& - \frac{1}{4 \pi}\int _{\partial \Sigma_\infty }
K^\beta {h_\mu}^{[\mu;\nu]} \, \varepsilon_{\beta \nu }
\, \varepsilon_{\alpha \mu }
\nonumber \\
   &=& \frac{1}{8 \pi}\int _{\partial \Sigma_\infty}
({h_\mu}^{\mu;\nu}-{h_\mu}^{\nu;\mu}) K^\beta \,
\varepsilon _{\beta \nu \alpha \mu }
\nonumber \\
   &=& - \delta M \; ,
   \label{wald}
\end{eqnarray}
where the orientation of $\varepsilon _{\beta \nu \alpha \mu }$ is
chosen so that $\varepsilon_{\beta \nu \alpha \mu }
= -6 \, \varepsilon_{[\beta \nu }\varepsilon_{\alpha \mu]}$ and
$\varepsilon _{\alpha \mu }$ is the volume element of the boundary
at infinity, and we have applied the result
$\frac{1}{8 \pi}\int _{\partial \Sigma_\infty}
({h_\mu}^{\mu;\nu}-{h_\mu}^{\nu;\mu})K^\beta \,
\varepsilon _{\beta \nu \alpha \mu } = - \delta M$
in the final step \cite{Wald1984}.

Eq.~(\ref{wald}) allows us to transform Eq.~(\ref{uuu}) into
\begin{equation}
   \delta M = -\delta M + \frac{1}{4 \pi}\int _{\partial \Sigma_\text{in} }
( K^\beta {h_\mu}^{[\mu;\nu]} - K^\nu {h_\mu}^{[\mu;\beta]} )
\hat N_\nu \, \hat T_\beta \,
\sqrt{|\gamma^{({\partial \Sigma_\text{in} })} |} \, d^2y
+ \frac{1}{4 \pi} \int _{\partial \Sigma_\text{in} }
\delta \kappa  \; dA
+ \frac{1}{4 \pi} \int _{\partial\Sigma_\text{in} } \kappa  \; \delta (dA).
   \label{differential 1}
\end{equation}
Or equivalently,
\begin{equation}
   \delta M = \frac{1}{8 \pi}\int _{\partial 
\Sigma_\text{in}}  \frac{1}{2}({h_\mu}^{\mu;\nu}-{h_\mu}^{\nu;\mu}) 
\hat N_\nu \, \hat T_\beta  K^\beta \, dA
+ \frac{1}{8 \pi} \int _{\partial\Sigma_\text{in} } 
\delta \kappa  \; dA 
+ \frac{1}{8 \pi} \int _{\partial\Sigma_\text{in} } \kappa  \; \delta (dA) \;,
   \label{differential 2}
\end{equation}
where we have used $K^\nu \hat N_\nu = {\cal N} \hat T^\nu \hat N_\nu=0$,
which follows since $\hat T^\mu$ is normal to
$\Sigma$ and $\hat N ^\mu$ lies in $\Sigma$.

For the first law to be true, we would require that the first two
boundary integrals of Eq.~(\ref{differential 2}) exactly cancel. In order
to further simplify these terms we start by considering the diffeomorphic
changes in more detail.

\vskip 0.2in

\end{widetext}

\subsection{Diffeomorphic conditions}
As already discussed, we assume
\begin{equation}
    \delta K^\mu=0, \qquad\qquad   \delta K_\mu=h_{\mu \nu} K^\nu .
    \label{Killing}
\end{equation}
Recall that by ``gauge'' freedom the sets of coordinates of $\Sigma$
and $\partial \Sigma$ are unchanged by the diffeomorphism, so without
loss of generality we may take \cite{Bardeen1973} 
\begin{equation}
    \delta (dx^\mu) = 0 \; , \forall dx^\mu \; \text{in} \; \Sigma.
\end{equation}

Because $K_\mu dx^\mu =  {\cal N} \hat T_\mu dx^\mu = 0$ for all
$dx^\mu$ in $\Sigma$, we have $\delta K_\mu \parallel K_\mu$, so
\begin{equation}
    \delta K_\mu = k_0 K_\mu.
    \label{kk1}
\end{equation}
Comparing Eq.~(\ref{kk1}) with Eq.~(\ref{Killing}), one finds 
\begin{equation}
   h_{\mu\nu} K^\nu = k_0 K_\mu
    \label{k0}
\end{equation}
everywhere. Then contracting $\hat T^\mu $ on both sides of this
equation yields
\begin{eqnarray}
k_0 = -h_{\mu\nu} \hat T^\mu \hat T^\nu.
    \label{k1}
\end{eqnarray}
(In other words, $k_0=-h_{\hat T \hat T}$ in the tetrad basis.)

Similarly, since $\hat T_\mu dx^\mu = 0$ for all $dx^\mu$ in
$\Sigma$, we have $\delta \hat T_\mu \parallel \hat T_\mu$, so
\begin{equation}
    \delta \hat T_\mu = \frac{k_1}{2} \hat T_\mu 
\end{equation}
(the factor of $\frac{1}{2}$ is for later convenience).
To get an expression for $k_1$, we calculate the variation of
$g^{\mu \nu} \hat T_\mu \hat T_\nu = -1$.
\begin{eqnarray}
(\delta g^{\mu \nu}) \hat T_\mu \hat T_\nu + 2 g^{\mu \nu}
\hat T_\mu (\delta \hat T_\nu ) &=& 0
\nonumber \\
\Rightarrow \;\;\;\; (- h^{\mu \nu}) \hat T_\mu \hat T_\nu + 2 g^{\mu \nu}
\hat T_\mu (\frac{k_1}{2} \hat T_\nu ) &=& 0 ,
\end{eqnarray}
hence
\begin{equation}
k_1 = - h_{\mu\nu} \hat T^\mu \hat T^\nu = k_0.
\label{eq35}
\end{equation}

For $\hat T^\mu$, we find
\begin{eqnarray}
\delta \hat T^\mu &=& \delta (g^{\mu \nu} \hat T_\nu) \nonumber \\
&=& -h^{\mu \nu} \hat T_\nu + \frac{1}{2} k_1 \hat T^\mu  \nonumber \\
&=& - g^{\mu \lambda} h_{\lambda \nu} \hat T^\nu
+ \frac{1}{2} k_1 \hat T^\mu \nonumber \\
&=& - g^{\mu \lambda} h_{\lambda \nu} K^\nu \hat T^t
+ \frac{1}{2} k_1 \hat T^\mu \nonumber \\
&=& - g^{\mu \lambda} k_1 K_\lambda \hat T^t
+ \frac{1}{2} k_1 T^\mu \nonumber \\
&=&  - k_1 \hat T^\mu + \frac{1}{2} k_1 \hat T^\mu \nonumber \\
&=&  - \frac{1}{2} k_1 \hat T^\mu ,
\label{eq36}
\end{eqnarray}
where we have used Eq.~(\ref{k0}) in the fourth line.

Again since, $dx^\mu \hat N_\mu = 0$ for all $dx^\mu$ in
$\partial \Sigma$, combined with $\hat T^\mu \hat N_\mu=0$, we find $\delta \hat N_\mu \parallel \hat N_\mu$, and so we write
\begin{equation}
    \delta \hat N_\mu = \frac{1}{2} k_2 \hat N_\mu .
\end{equation}
(The factor $\frac{1}{2}$ is introduced for later convenience.)
In the same way as for Eq.~(\ref{k1}) we find
\begin{eqnarray}
k_2 = h_{\mu\nu} \hat N^\mu \hat N^\nu, \qquad\qquad 
h_{\lambda \nu}  \hat T^\lambda \hat N^\nu =0 .
\label{k2}
\end{eqnarray}

Let us now introduce the whole tetrad basis $\{ \hat T^\mu, \hat N^\mu, \hat U^\mu, \hat V^\mu \}$; recall $\hat T^\mu$ is normal to $\Sigma$, $\hat N^\mu$ is in $\Sigma$ but normal to $\partial \Sigma$, and $\hat U^\mu, \hat V^\mu$ lie in $\partial \Sigma$. The projector onto $\partial \Sigma$ is defined
\begin{equation}
    P^{\mu\nu}\equiv (\hat U \otimes  \hat U + \hat V \otimes  \hat V)^{\mu\nu} = \hat U^\mu  \hat U^\nu + \hat V^\mu \hat V^\nu.
\end{equation}
Similarly
\begin{equation}
    g^{\mu\nu} = - \hat T^\mu  \hat T^\nu + \hat N^\mu \hat N^\nu + P^{\mu\nu} .
    \label{tetradee}
\end{equation}
Now tangent vectors in $\partial \Sigma$ are contained in $\text{span} \{ \hat U^\mu, \hat V^\mu \}$ and since the coordinates of 
$\partial \Sigma$ are preserved under the diffeomorphism, $\delta U^{\mu}$ and $ \delta V^{\mu}$ must also be contained in $\text{span} \{ \hat U^\mu, \hat V^\mu \}$. By the same reasoning, $\delta P^{\mu\nu} \in \text{span} \{ \hat U \otimes  \hat U, \hat U \otimes  \hat V, \hat V \otimes  \hat U, \hat V \otimes  \hat V \}$.

Let us now use the tetrad decomposition for $\hat N^\mu$ to find
\begin{eqnarray}
\delta \hat N^\mu &=& \delta (g^{\mu \nu} \hat N_\nu) 
\nonumber \\
&=& -h^{\mu \nu} \hat N_\nu + \frac{1}{2} k_2 \hat N^\mu
\nonumber \\
&=& - g^{\mu \lambda} h_{\lambda \nu} \hat N^\nu
+ \frac{1}{2} k_2 \hat N^\mu
\nonumber \\
&=& - g^{\mu \lambda} (-\delta (\hat T_\lambda \hat T_\nu)
+ \delta (\hat N_\lambda \hat N_\nu)
+ \delta P_{\lambda \nu}) \hat N^\nu + \frac{1}{2} k_2 \hat N^\mu
\nonumber \\
&=& - \frac{1}{2} k_2 \hat N^\mu - g^{\mu \lambda}
\delta P_{\lambda \nu} \hat N^\nu
\nonumber \\
&=& - \frac{1}{2} k_2 \hat N^\mu   .
\end{eqnarray}
Note in the fifth line we use $ \delta P_{\lambda \nu} \hat N^\nu =0$
since $\delta P_{\mu\nu} \in \text{span} \{ \hat U \otimes  \hat U, \hat U \otimes  \hat V, \hat V \otimes  \hat U, \hat V \otimes  \hat V \}$.

Further, since $\delta U^{\mu}, \delta V^{\mu} \in \text{span} \{ \hat U^\mu, \hat V^\mu  \}$, we may explicitly write them as
\begin{eqnarray}
    \delta \hat U^\mu = -\frac{1}{2} k_3 \hat U^\mu -\frac{1}{2} k_4 \hat V^\mu,  \;\;
    \delta \hat V^\mu = -\frac{1}{2} k_5 \hat U^\mu -\frac{1}{2} k_6 \hat V^\mu .
    \nonumber \\
\end{eqnarray}
By considering $\delta(g_{\mu\nu} \hat U^\mu \hat U^\nu) = 0$, $\delta(g_{\mu\nu} \hat U^\mu \hat V^\nu) = 0$ and $\delta(g_{\mu\nu} \hat V^\mu \hat V^\nu) = 0$, it is easy to show that
\begin{eqnarray}
    &&k_3=h_{\mu \nu} \hat U^\mu \hat U^\nu \nonumber \\
    &&k_6=h_{\mu \nu} \hat V^\mu \hat V^\nu \nonumber \\
    &&k_4+k_5 = 2 h_{\mu \nu} \hat U^\mu \hat V^\nu .
    \label{k3k6}
\end{eqnarray}
Then by considering $\delta (\hat U_\mu \hat U^\mu)=0$, $\delta (\hat U_\mu \hat V^\mu)=0$, $\delta (\hat V_\mu \hat V^\mu)=0$ and $\delta (\hat V_\mu \hat U^\mu)=0$, one finds
\begin{eqnarray}
    \delta \hat U_\mu = \frac{1}{2} k_3 \hat U_\mu +\frac{1}{2} k_5 \hat V_\mu,  \;\;\;
    \delta \hat V_\mu = \frac{1}{2} k_4 \hat U_\mu +\frac{1}{2} k_6 \hat V_\mu .
    \nonumber \\
\end{eqnarray}
Hence,  $\delta P^{\mu \nu}$ may be explicitly computed to be
\begin{eqnarray}
\delta P^{\mu \nu} &=& - k_3 \hat U^\mu
\hat U^\nu - k_6 \hat V^\mu \hat V^\nu\nonumber \\
&&-\frac{1}{2} (k_4+k_5) (\hat U^\mu \hat V^\nu + \hat U^\nu \hat V^\mu ) 
\nonumber \\
&=& -\frac{1}{2}( k_3+k_6) (\hat U^\mu \hat U^\nu + \hat V^\mu \hat V^\nu )\nonumber\\
&& -\frac{1}{2} ( k_3-k_6) (\hat U^\mu \hat U^\nu - \hat V^\mu \hat V^\nu ) 
\nonumber \\
&&-\frac{1}{2} (k_4+k_5) (\hat U^\mu \hat V^\nu + \hat U^\nu \hat V^\mu )
\nonumber \\
&=& -\frac{1}{2}( k_3+k_6) P^{\mu \nu} -\frac{1}{2} ( k_3-k_6) (\hat U^\mu \hat U^\nu
- \hat V^\mu \hat V^\nu ) 
\nonumber \\
&&-\frac{1}{2} (k_4+k_5) (\hat U^\mu \hat V^\nu + \hat U^\nu \hat V^\mu ) ,
\label{projection}
\end{eqnarray}
Similarly,
\begin{eqnarray}
\delta P_{\mu \nu}  &=& k_3 \hat U_\mu \hat U_\nu + k_6 \hat V_\mu \hat V_\nu \nonumber \\
&& + \frac{1}{2} (k_4+k_5) (\hat U_\mu \hat V_\nu + \hat U_\nu \hat V_\mu ) 
\nonumber \\
&=& \frac{1}{2}( k_3+k_6) P_{\mu \nu} + \frac{1}{2} ( k_3-k_6) (\hat U_\mu \hat U_\nu
- \hat V_\mu \hat V_\nu ) 
\nonumber \\
&& + \frac{1}{2} (k_4+k_5) (\hat U_\mu \hat V_\nu + \hat U_\nu \hat V_\mu ).
\label{projection2}
\end{eqnarray}

So the key diffeomorphic conditions may be summarized as
\begin{eqnarray}
&&  \delta K^{\mu }=0, \;\;\;\;\;\;\;\;\;\;\;\;\;\;
 \delta K_\mu=h_{\mu \nu} K^\nu = k_1 K_\mu
\nonumber \\
&& \delta \hat T^\mu=  - \frac{1}{2} k_1 \hat T^\mu,
\;\;\;\,  \delta \hat T_\mu = \frac{1}{2} k_1 \hat T_\mu 
\nonumber \\
&& \delta \hat N^\mu = - \frac{1}{2} k_2 \hat N^\mu,
\;\; \delta \hat N_\mu = \frac{1}{2}  k_2 \hat N_\mu
\nonumber \\
&& \delta \hat T^t = - \frac{1}{2} k_1 \hat T^t, 
\;\;\;\;\; \hat T^\mu h_{\mu \nu} \hat N^\nu = 0 
\nonumber \\
&& \delta P^{\mu \nu} = \text{Eq.}(\ref{projection}) , \;\;\;\, \delta P_{\mu \nu} = \text{Eq.}(\ref{projection2}) .
\label{alll}
\end{eqnarray}
where $k_1 = -h_{\mu\nu} \hat T^\mu \hat T^\nu$ and
$k_2 = h_{\mu\nu} \hat N^\mu \hat N^\nu$.

\def \blah {

\subsection*{Differential inner boundary term}

As a preliminary, recall that the surface gravity extended to other
surfaces is given by
$\kappa \equiv {K^\mu}_{;\nu} \hat T^\nu \hat N_\mu$ from
Eq.~(\ref{def}). Next, we extend $\hat N_\mu$ as a vector field
inside $\Sigma$ away from $\partial \Sigma$, consistent with
$\hat T^\mu \hat N_\mu = 0$ and $\hat N^\mu \hat N_\mu = 1$. 
Thus, the extended surface gravity becomes
\begin{equation}
    \kappa \equiv K_{\mu ;\nu} \hat T^\nu \hat N^\mu
= K_{\mu , \nu} \hat T^\nu \hat N^\mu -
\Gamma_{\mu \nu}^\lambda K_\lambda \hat{T}^\nu \hat N^\mu .
\end{equation}
The diffeomorphic variation of $\kappa$ may therefore be computed as
\begin{widetext}
\begin{eqnarray}
    \delta \kappa &=& (\delta K_\mu)_{ , \nu}
\hat T^\nu \hat N^\mu + K_{\mu , \nu}
\delta (\hat T^\nu \hat N^\mu) -
(\delta \Gamma_{\mu \nu}^\lambda) K_\lambda
 \hat{T}^\nu \hat N^\mu - \Gamma_{\mu \nu}^\lambda
\delta (K_\lambda  \hat{T}^\nu \hat N^\mu ) \nonumber \\
    &=& k_1  K_{\mu , \nu} \hat T^\nu \hat N^\mu -
(\frac{1}{2}k_1 +\frac{1}{2}k_2) K_{\mu , \nu}
\hat T^\nu \hat N^\mu - (\delta \Gamma_{\mu \nu}^\lambda)
K_\lambda  \hat{T}^\nu \hat N^\mu - (k_1-\frac{1}{2}k_1
- \frac{1}{2}k_2) \Gamma_{\mu \nu}^\lambda K_\lambda 
\hat{T}^\nu \hat N^\mu \nonumber \\
    &=& (k_1-\frac{1}{2}k_1 - \frac{1}{2}k_2 ) ( K_{\mu , \nu}
\hat T^\nu \hat N^\mu - \Gamma_{\mu \nu}^\lambda K_\lambda 
\hat{T}^\nu \hat N^\mu ) - (\delta \Gamma_{\mu \nu}^\lambda)
K_\lambda  \hat{T}^\nu \hat N^\mu \nonumber \\
    &=& (\frac{1}{2} k_1 - \frac{1}{2} k_2) \, \kappa
- \frac{1}{2} g^{\lambda \rho } ( h_{\mu \rho ;\nu }
+ h_{\nu \rho ;\mu }-h_{\mu \nu ;\rho } )
 K_\lambda  \hat{T}^\nu \hat N^\mu \nonumber \\
    &=& (\frac{1}{2} k_1 - \frac{1}{2} k_2) \, \kappa
- \frac{1}{2} ( h_{\mu \rho ;\nu }
+ h_{\nu \rho ;\mu }-h_{\mu \nu ;\rho } )  K^\rho
\hat{T}^\nu \hat N^\mu \nonumber \\
    &=& (\frac{1}{2} k_1 - \frac{1}{2} k_2) \, \kappa
- \frac{1}{2} ( h_{\mu \rho ;\nu } + h_{\nu \rho ;\mu }
-h_{\mu \nu ;\rho } ) \frac{1}{\hat T^t} \hat T^\rho
\hat{T}^\nu \hat N^\mu
    \nonumber \\
    &=& (\frac{1}{2} k_1 - \frac{1}{2} k_2) \,
\kappa - \frac{1}{2}  h_{\nu \rho ;\mu }  \hat T^\rho
\hat{T}^\nu \hat N^\mu {\cal N} \nonumber \\
    &=& (\frac{1}{2} k_1 - \frac{1}{2} k_2) \, \kappa
- \frac{1}{2}  h_{\nu \rho ;\mu } ( \hat N^\rho \hat{N}^\nu
+ P^{\nu \rho} - g^{\nu \rho})\hat N^\mu {\cal N} \nonumber \\
    &=& (\frac{1}{2} k_1 - \frac{1}{2} k_2) \, \kappa
+ \frac{1}{2} {h^\nu}_{  \nu ;\mu } \hat N^\mu {\cal N}
 -  \frac{1}{2} h_{\nu \rho ;\mu } \hat N^\rho \hat{N}^\nu
\hat N^\mu {\cal N} - \frac{1}{2}   h_{\nu \rho ;\mu }
P^{\nu \rho} \hat N^\mu {\cal N} \nonumber \\
    &=& (\frac{1}{2} k_1 - \frac{1}{2} k_2) \, \kappa
+ \frac{1}{2} {h^\nu}_{  \nu ;\mu } \hat N^\mu {\cal N}
-  \frac{1}{2} h_{\nu \rho ;\mu } \hat N^\rho ( g^{\mu\nu}
+ \hat{T}^\nu \hat T^\mu - P^{\mu\nu} ) {\cal N} -  \frac{1}{2} 
 h_{\nu \rho ;\mu } P^{\nu \rho} \hat N^\mu {\cal N} \nonumber \\
    &=& (\frac{1}{2} k_1 - \frac{1}{2} k_2) \, \kappa
+ \frac{1}{2} {h^\nu}_{  \nu ;\mu } \hat N^\mu {\cal N}
 -  \frac{1}{2} {h^\mu}_{\rho ;\mu } \hat N^\rho  {\cal N}
-  \frac{1}{2} h_{\nu \rho ;\mu } \hat N^\rho ( \hat{T}^\nu
\hat T^\mu - P^{\mu\nu} ) {\cal N} -  \frac{1}{2} 
 h_{\nu \rho ;\mu } P^{\nu \rho} \hat N^\mu {\cal N}
    \nonumber \\
    &=& (\frac{1}{2} k_1 - \frac{1}{2} k_2) \, \kappa +
\frac{1}{2} {h^\nu}_{  \nu ;\mu } \hat N^\mu {\cal N}
 -  \frac{1}{2} {h^\mu}_{\rho ;\mu } \hat N^\rho  {\cal N}
-  \frac{1}{2} h_{\nu \rho ;\mu } (\hat N^\rho \hat{T}^\nu
\hat T^\mu - \hat N^\rho P^{\mu\nu} + P^{\nu \rho}
\hat N^\mu ) {\cal N} \; ,
    \label{vark}
\end{eqnarray}
where we have used $\delta \Gamma_{\mu \nu}^\lambda = \frac{1}{2}
g^{\lambda \rho } ( h_{\mu \rho ;\nu }
+ h_{\nu \rho ;\mu }-h_{\mu \nu ;\rho } )$ in the third
line \cite{palatini1919}.

Since $K^\mu \hat N_\mu=0 $, then $( K^\mu \hat N_\mu )_{;\nu} = 0 $, and so 
\begin{equation}
    {K^\mu}_{;\nu} \hat N_\mu = - K^\mu \hat N_{\mu;\nu} .
    \label{nnnmu}
\end{equation} 
Further, 
\begin{eqnarray}
   h_{\mu \nu} = \delta g_{\mu \nu } &=& \delta (-\hat T_\mu
\hat T_\nu + \hat N_\mu \hat N_\nu + P_{\mu \nu}) \nonumber \\
    &=& - k_1 \hat T_\mu \hat T_\nu + k_2 \hat N_\mu \hat N_\nu
+ \delta P_{\mu \nu} \; .
    \label{hhmunu}
\end{eqnarray}
We now use Eqs.~(\ref{nnnmu}) and (\ref{hhmunu}) to simplify the
last term of Eq.(\ref{vark}) as
\begin{eqnarray}
    && \frac{1}{2} h_{\nu \rho ;\mu } (\hat N^\rho \hat{T}^\nu
\hat T^\mu - \hat N^\rho P^{\mu\nu} + P^{\nu \rho} \hat N^\mu )
{\cal N} \nonumber \\ 
    &=&  \frac{1}{2} (- k_1 \hat T_\nu \hat T_\rho
+ k_2 \hat N_\nu \hat N_\rho + \delta P_{\nu \rho} )_{;\mu }
(\hat N^\rho \hat{T}^\nu \hat T^\mu - \hat N^\rho P^{\mu\nu}
+ P^{\nu \rho} \hat N^\mu ) {\cal N} \nonumber \\
    &=&  \frac{1}{2} k_1 \hat T_{\rho;\mu} \hat N^\rho \hat T^\mu
 {\cal N} + \frac{1}{2} k_2 \hat N_{\nu ; \mu} \hat T^\nu
\hat T^\mu  {\cal N} - \frac{1}{2} k_2 \hat N_{\nu ; \mu}
P^{\mu\nu} {\cal N} + \frac{1}{2} (\delta P_{\nu \rho} )_{;\mu }
(\hat N^\rho \hat{T}^\nu \hat T^\mu - \hat N^\rho P^{\mu\nu}
+ P^{\nu \rho} \hat N^\mu ) {\cal N} \nonumber \\
    &=&  \frac{1}{2} k_1 K_{\rho;\mu} \hat N^\rho \hat T^\mu
+ \frac{1}{2} k_2 \hat N_{\nu ; \mu} K^\nu \hat T^\mu
- \frac{1}{2} k_2 \hat N_{\nu ; \mu} P^{\mu\nu} {\cal N}
+ \frac{1}{2} (\delta P_{\nu \rho} )_{;\mu } (\hat N^\rho
\hat{T}^\nu \hat T^\mu - \hat N^\rho P^{\mu\nu} + P^{\nu \rho}
\hat N^\mu ) {\cal N}
    \nonumber \\
    &=&  \frac{1}{2} k_1 \, \kappa - \frac{1}{2} k_2 \,
\kappa - \frac{1}{2} k_2 \hat N_{\mu ; \nu} P^{\mu\nu} {\cal N}
+ \frac{1}{2} (\delta P_{\nu \rho} )_{;\mu } (\hat N^\rho
\hat{T}^\nu \hat T^\mu - \hat N^\rho P^{\mu\nu}
+ P^{\nu \rho} \hat N^\mu ) {\cal N} \; .
    \label{vark2}
\end{eqnarray}

Put Eq.(\ref{vark2}) back into Eq.(\ref{vark}) to yield 
\begin{eqnarray}
    \delta \kappa &=& \frac{1}{2} {h^\nu}_{  \nu ;\mu }
\hat N^\mu {\cal N}  -  \frac{1}{2} {h^\mu}_{\rho ;\mu }
\hat N^\rho  {\cal N} + \frac{1}{2} k_2 \hat N_{\mu ; \nu}
P^{\mu\nu} {\cal N} - \frac{1}{2} (\delta P_{\nu \rho} )_{;\mu }
(\hat N^\rho \hat{T}^\nu \hat T^\mu - \hat N^\rho P^{\mu\nu}
+ P^{\nu \rho} \hat N^\mu ) {\cal N} \; .
    \label{vark3}
\end{eqnarray}
Then Eq.~(\ref{projection2}) helps us transform the last term
of Eq.~(\ref{vark3}) into
\begin{eqnarray}
    && \frac{1}{2} (\delta P_{\nu \rho} )_{;\mu } (\hat N^\rho
\hat{T}^\nu \hat T^\mu - \hat N^\rho P^{\mu\nu}
+ P^{\nu \rho} \hat N^\mu ) {\cal N} \nonumber \\
    &=& \frac{1}{2} \Bigl( (\hat N^\rho \delta
P_{\nu \rho} )_{;\mu } \hat{T}^\nu \hat T^\mu
- (\hat N^\rho)_{;\mu } \hat{T}^\nu \hat T^\mu
\delta P_{\nu \rho}  - (\hat N^\rho \delta P_{\nu \rho} )_{;\mu }
P^{\mu\nu} + (\hat N^\rho )_{;\mu } P^{\mu\nu}
\delta P_{\nu \rho}  + (\delta P_{\nu \rho} )_{;\mu }
P^{\nu \rho} \hat N^\mu \Bigl ) {\cal N} \nonumber \\
    &=& \frac{1}{2} \Bigl( (\hat N^\rho )_{;\mu } P^{\mu\nu}
\delta P_{\nu \rho} + (\delta P_{\nu \rho} )_{;\mu }
\hat N^\mu P^{\nu \rho} \Bigl ) {\cal N} \nonumber \\
    &=&  \frac{1}{2} \biggl( \hat N_{\rho ;\mu }
\Bigl (\frac{1}{2}( k_3+k_6) P^{\rho \mu } + \frac{1}{2} ( k_3-k_6)
(\hat U^\rho \hat U^\mu - \hat V^\rho \hat V^\mu )
+ \frac{1}{2} (k_4+k_5) (\hat U^\rho \hat V^\mu
+ \hat U^\mu \hat V^\rho ) \Bigl ) + (\delta P_{\nu \rho} )_{;\mu }
\hat N^\mu P^{\nu \rho} \biggl ) {\cal N} . \nonumber \\
    \label{vark4}
\end{eqnarray}
To simplify further, we firstly show that
$ P^{\mu \nu} (\delta P_{\mu \nu })_{;\rho } \hat N^\rho
=  (k_3+k_6)_ {;\rho } \hat N^\rho$ :
\begin{eqnarray}
  P^{\mu \nu} (\delta P_{\mu \nu })_{;\rho } \hat N^\rho 
  &=&  P^{\mu \nu} \Bigl (  k_3 \hat U_\mu \hat U_\nu +
k_6 \hat V_\mu \hat V_\nu + \frac{1}{2}(k_4+k_5)
(\hat U_\mu \hat V_\nu + \hat U_\nu \hat V_\mu ) \Bigl )_{;\rho } \hat N^\rho  
  \nonumber \\
   &=& {k_3}_ {;\rho } \hat N^\rho + 2 k_3 \hat U_{\mu ; \rho }
\hat N^\rho \hat U^\mu + {k_6}_{;\rho } \hat N^\rho + 2 k_6 \,
\hat V_{\mu ; \rho } \hat N^\rho \hat V^\mu + \frac{1}{2} (k_4+k_5)
 P^{\mu \nu} (\hat U_\mu \hat V_\nu + \hat U_\nu
\hat V_\mu )_{;\rho } \hat N^\rho 
   \nonumber \\
   &=& {k_3}_ {;\rho } \hat N^\rho  + {k_6}_{;\rho } \hat N^\rho
+(k_4+k_5) (\hat U_{\mu ;\rho } \hat N^\rho \hat V^\mu +
\hat V_{\mu ;\rho } \hat N^\rho \hat U^\mu)
   \nonumber \\
   &=& {k_3}_ {;\rho } \hat N^\rho  + {k_6}_{;\rho } \hat N^\rho
+ (k_4+k_5) (\hat U_\mu \hat V ^\mu)_{ ;\rho } \hat N^\rho
   \nonumber \\
   &=& (k_3+k_6)_ {, \rho } \hat N^\rho  \;,
   \label{pmunu}
\end{eqnarray}
where we have used $ \hat U_{\mu ; \rho } \hat N^\rho \hat U^\mu
= \frac{1}{2} (\hat U_\mu \hat U^\mu)_{; \rho } \hat N^\rho  =0$
for both $\hat U^\mu$ and $\hat V^\mu$ in the second line. 

Now substitute Eq.~(\ref{pmunu}) into Eq.~(\ref{vark4}) and then
replace the last term in Eq.~(\ref{vark3}) by Eq.~(\ref{vark4}), we find 
\begin{eqnarray}
    \delta \kappa &=& \frac{1}{2} {h^\nu}_{  \nu ;\mu } \hat N^\mu {\cal N}  -  \frac{1}{2} {h^\mu}_{\rho ;\mu } \hat N^\rho  {\cal N} + \frac{1}{2} k_2 \hat N_{\mu ; \nu} P^{\mu\nu} {\cal N} \nonumber \\ 
    &&- \frac{1}{2} \biggl( \hat N_{\rho ;\mu }  \Bigl ( \frac{1}{2}( k_3+k_6) P^{\rho \mu } + \frac{1}{2} ( k_3-k_6) (\hat U^\rho \hat U^\mu - \hat V^\rho \hat V^\mu ) + \frac{1}{2} (k_4+k_5) (\hat U^\rho \hat V^\mu + \hat U^\mu \hat V^\rho ) \Bigl) + (k_3+k_6)_ {, \rho } \hat N^\rho \biggl ) {\cal N} \nonumber \\
     &=& \frac{1}{2} {h^\nu}_{  \nu ;\mu } \hat N^\mu {\cal N} - \frac{1}{2} {h^\mu}_{\rho ;\mu } \hat N^\rho  {\cal N} \nonumber \\ 
    &&- \frac{1}{2} \biggl( \hat N_{\rho ;\mu }  \Bigl(\frac{1}{2}( k_3+k_6 - 2k_2) P^{\rho \mu } + \frac{1}{2} ( k_3-k_6) (\hat U^\rho \hat U^\mu - \hat V^\rho \hat V^\mu ) + \frac{1}{2} (k_4+k_5) (\hat U^\rho \hat V^\mu + \hat U^\mu \hat V^\rho ) \Bigl) + (k_3+k_6)_ {, \rho } \hat N^\rho \biggl ) {\cal N}  \; . \nonumber \\
    \label{vark5}
\end{eqnarray}

Substituting Eq.~(\ref{vark5}) into Eq.~(\ref{differential 2}) and recalling that $\hat T_\beta K^\beta= - {\cal N}$, we finally find 
\begin{eqnarray}
    \, \delta M &=& -\frac{1}{16 \pi} \int _{\partial\Sigma_\text{HS} }
    \biggl( \hat N_{\rho ;\mu }  \Bigl ( \frac{1}{2}( k_3+k_6 - 2k_2) P^{\rho \mu } + \frac{1}{2} ( k_3-k_6) (\hat U^\rho \hat U^\mu - \hat V^\rho \hat V^\mu ) + \frac{1}{2} (k_4+k_5) (\hat U^\rho \hat V^\mu + \hat U^\mu \hat V^\rho ) \Bigl ) \nonumber \\
    && + (k_3+k_6)_ {, \rho } \hat N^\rho \biggl ) {\cal N} \; dA + \frac{1}{8 \pi} \int _{\partial\Sigma_\text{HS} } \kappa \; \delta (dA).
    \label{HS 1st law}
\end{eqnarray}
Geometrically, the extrinsic curvature of the inner boundary is just
given by
\begin{eqnarray}
    K \equiv \hat N_{\rho ;\mu } P^{\rho \mu }
\end{eqnarray}
Similarly, we define the local shears as
\begin{eqnarray}
    \sigma _+  \equiv  \hat N_{\rho ;\mu } (\hat U^\rho \hat U^\mu
 - \hat V^\rho \hat V^\mu ) \; ; \;\;\;\; \sigma _\times
\equiv \hat N_{\rho ;\mu } (\hat U^\rho \hat V^\mu
 + \hat U^\mu \hat V^\rho ) \;.
\end{eqnarray}
Then Eq.(\ref{HS 1st law}) becomes 
\begin{equation}
    \, \delta M = -\frac{1}{32 \pi} \int _{\partial\Sigma_\text{HS} }
    \!\!\Bigl(  ( k_3+k_6 - 2k_2) K
+  ( k_3-k_6) \sigma _+
+  (k_4+k_5) \sigma _\times
+ 2(k_3+k_6)_ {, \rho } \hat N^\rho \Bigr)
{\cal N} \, dA 
    + \frac{1}{8 \pi} \int _{\partial\Sigma_\text{HS} }
\!\!\!\kappa \; \delta (dA). 
\label{ManResult}
\end{equation}
Again, noting that the matter inside the inner boundary is not
necessary to be a black hole.

\vskip 0.1in
\end{widetext}

}

From Eqs.~(\ref{k1}), (\ref{k2}) and (\ref{k3k6}), we know that
in the tetrad basis that
\begin{equation}
    h_{\hat \mu \hat \nu}=
\begin{pmatrix}
-k_1 & 0 & 0 & 0\\ 
0 & k_2 & 0 & 0\\ 
0 & 0 & k_3 & \frac{k_4+k_5}{2}\\ 
0 & 0 & \frac{k_4+k_5}{2} & k_6
\end{pmatrix},
\end{equation}
i.e., $h_{\hat T \hat T} = -k_1, h_{\hat N \hat N} = k_2 $
etc.. So $k_2, k_3,\ldots, k_6$ are independent functions from each other.

\def\blah{
,
and thus the first surface integral of Eq.~(\ref{ManResult}) vanishes
only when either
${\cal N} = 0$, i.e., the horizon for a simple static spacetime (if it
contains a black hole); or each term $ K$, $\sigma _+$,
$\sigma _\times$ etc. must vanish at every point on
$\partial\Sigma_{\text{HS}}$. It is worth pointing out that for
holographic screens of order a Planck length outside a black hole's
horizon (often called stretched horizons), 
${\cal N}^2 \equiv O(\frac{E_{\text{Planck}}}{M})$,
i.e., the lapse function still approximately vanishes. The first law
will hold up to corrections smaller than a Planck energy.
}

\subsection{Reduction to the first law}


Since $K^\mu \hat N_\mu=0 $, then $( K^\mu \hat N_\mu )_{;\nu} = 0 $, and so 
\begin{equation}
    {K^\mu}_{;\nu} \hat N_\mu = - K^\mu \hat N_{\mu;\nu} .
    \label{nnnmu}
\end{equation} 
We then consider the expansion of null normal congruences on the inner
boundary which may be written as
\begin{eqnarray}
   \theta ^{(l)} &=& P^{\mu \nu} l_{\mu_ ;\nu} \nonumber \\
   &=& P^{\mu \nu} (\hat T_\mu + \hat N_\mu)_{;\nu} \nonumber \\
   &=& P^{\mu \nu} (\hat T^t K_\mu)_{ ;\nu} +   P^{\mu \nu}
\hat N_{\mu;\nu} \nonumber \\
   &=& \hat T^t P^{\mu \nu} K_{\mu;\nu}
+ P^{\mu \nu} K_\mu (\hat T^t)_{ ;\nu}
+   P^{\mu \nu} \hat N_{\mu;\nu} \nonumber \\
   &=& P^{\mu \nu} \hat N_{\mu;\nu} \nonumber \\
   &=& ( g^{\mu \nu}
+ \hat T^\mu \hat T^\nu  -
\hat N^\mu \hat N^\nu ) \hat N_{\mu ;\nu} \nonumber \\
   &=& {\hat N^\mu}_{\;\;;\mu} - K_{\mu ; \nu}
\hat T^\nu \hat N^\mu \hat T^t 
   \nonumber \\
   &=& {\hat N^\mu}_{\;\;;\mu} - \kappa \, \hat T^t
   \nonumber \\
   &=& \frac{1}{\sqrt{-g}} (\sqrt{-g} \hat N^\mu)_{, \mu}
- \kappa \, \hat T^t
   \label{expansion}
\end{eqnarray}
where $ l _\mu = \hat T_\mu + \hat N_\mu$ is the outgoing null normal
vector of the inner boundary, and we have used Eq.~(\ref{tetradee}) in
the fifth line and Eq.~(\ref{nnnmu}) in the sixth line. Using this
relation we may express the variation of $\kappa \hat T^t$ as
\begin{widetext}
\begin{eqnarray}
    \delta \hat T^t \kappa + \hat T^t \delta \kappa &=&
\delta (\frac{1}{\sqrt{-g}} ) (\sqrt{-g}
\hat N^\mu)_{, \mu} +\frac{1}{\sqrt{-g}} (\delta\sqrt{-g}
\hat N^\mu)_{, \mu} +\frac{1}{\sqrt{-g}}
(\sqrt{-g} \delta \hat N^\mu)_{, \mu} -  \, \delta \theta ^{(l)}  
  \nonumber \\
    -\frac{1}{2} k_1 \hat T^t \kappa + \hat T^t \delta \kappa &=&
-\frac{1}{2} g^{\tau \nu} \delta g_{\tau \nu} \frac{1}{\sqrt{-g}}
 (\sqrt{-g} \hat N^\mu)_{, \mu} +\frac{1}{\sqrt{-g}}
(\frac{1}{2} \sqrt{-g} g^{\tau \nu}
\delta g_{\tau \nu} \hat N^\mu)_{, \mu} + 
( \delta \hat N^\mu)_{; \mu} -  \, \delta \theta ^{(l)}
    \nonumber \\
    &=& - \frac{1}{2} g^{\tau \nu} \delta g_{\tau \nu}
\frac{1}{\sqrt{-g}}  (\sqrt{-g} \hat N^\mu)_{, \mu}
+\frac{1}{2} g^{\tau \nu} \delta g_{\tau \nu}
\frac{1}{\sqrt{-g}} ( \sqrt{-g} \hat N^\mu)_{, \mu}
+\frac{1}{2} ( g^{\tau \nu} \delta g_{\tau \nu} )_{, \mu}
\hat N^\mu + ( \delta \hat N^\mu)_{; \mu} -  \, \delta \theta ^{(l)} 
    \nonumber \\
    &=& \frac{1}{2} ( {h_\nu }^\nu)_{, \mu} \hat N^\mu
+ ( \delta \hat N^\mu)_{; \mu} -  \, \delta \theta ^{(l)}  \; ,
\end{eqnarray}
where $\delta \sqrt{-g} = \frac{1}{2} \sqrt{-g} g^{\mu \nu}
\delta g_{\mu \nu}$.
Hence, the first term in Eq.~(\ref{differential 2}) may be written as
\begin{equation}
   \frac{1}{2} ( {h_\nu }^\nu)_{; \mu} \hat N^\mu 
= -\frac{1}{2} k_1 \hat T^t \kappa + \hat T^t \delta \kappa
- ( \delta \hat N^\mu)_{; \mu} + \, \delta \theta ^{(l)}
   \label{32}
\end{equation}

The second term ${h_\mu}^{\nu;\mu} \hat N_\nu$ in Eq.(\ref{differential 2})
can then be expressed as 
\begin{eqnarray}
     \frac{1}{2} { h_\mu}^{\nu ;\mu} \hat N_\nu  
    &=& \frac{1}{2} ({h_\mu}^\nu \hat N_\nu )^{;\mu}
- \frac{1}{2} {h_\mu}^\nu \hat N_\nu ^{\;\;\,;\mu}  
    = \frac{1}{2} (h_{\mu \nu  } \hat N^\nu )^{;\mu}
- \frac{1}{2} h^{\mu \nu} \hat N _{\nu;\mu} 
   = \frac{1}{2} (\delta g_{\mu \nu} \hat N^\nu )^{;\mu}
+ \frac{1}{2} \delta g^{\mu \nu } \hat N_{\mu;\nu} 
    \nonumber \\
     &=& \frac{1}{2} \bigl (\delta (g_{\mu \nu} \hat N^\nu) -  g_{\mu \nu} \delta \hat N^\nu \bigr)^{;\mu} + \frac{1}{2} \delta ( - \hat T^\mu
\hat T^\nu +\hat N^\mu \hat N^\nu + P^{\mu \nu } ) \hat N_{\mu;\nu}  
    \nonumber \\
    &=&\frac{1}{2} (\delta \hat N_\mu - \delta \hat N^\nu g_{\mu \nu  } )^{;\mu} + \frac{1}{2} ( k_1 \hat T^\mu \hat T^\nu - k_2 \hat N^\mu \hat N^\nu + \delta P^{\mu \nu } ) \hat N_{\mu;\nu}
    \nonumber \\
    &=&\frac{1}{2} (\frac{1}{2} k_2 \hat N_\mu )^{;\mu} -\frac{1}{2} (\delta \hat N^\mu)_{;\mu} + \frac{1}{2} k_1 \hat T^\mu \hat T^\nu \hat N_{\mu;\nu} + \frac{1}{2} \delta P^{\mu \nu} \hat N_{\mu;\nu}
   \nonumber \\
     &=&\frac{1}{2} (\frac{1}{2} k_2 \hat N^\mu )_{;\mu} -\frac{1}{2} (\delta \hat N^\mu)_{;\mu} - \frac{1}{2} k_1 \hat T^t \hat T^\nu \hat N_\mu {K^\mu}_{;\nu} + \frac{1}{2} \delta P^{\mu \nu} \hat N_{\mu;\nu}
   \nonumber \\
    &=& - (\delta \hat N^\mu)_{;\mu} - \frac{1}{2} k_1 \hat T^t \kappa + \frac{1}{2} \delta P^{\mu \nu} \hat N_{\mu;\nu} ,
\label{piupiu}
\end{eqnarray}
where we have used Eq.~(\ref{nnnmu}) in the fifth line, and
Eqs.~(\ref{def}) and (\ref{alll})  in the last step.

Next, consider the final term
$\frac{1}{2} \delta P^{\mu \nu} \hat N_{\mu ;\nu}$ in
Eq.~(\ref{piupiu}), using Eq.~(\ref{projection}) we have
\begin{eqnarray}
    \frac{1}{2} \delta P^{\mu \nu} \hat N_{\mu;\nu} 
    &=& 
-\frac{1}{4} \Big( ( k_3+k_6) P^{\mu \nu} + ( k_3-k_6) (\hat U^\mu \hat U^\nu
- \hat V^\mu \hat V^\nu ) + (k_4+k_5)
(\hat U^\mu \hat V^\nu + \hat U^\nu \hat V^\mu ) \Big )
\hat N_{\mu;\nu}
    \nonumber \\
    &=&  -\frac{1}{4} \Big( ( k_3+k_6) P^{\mu \nu}
+ ( k_3-k_6) (\hat U^\mu \hat U^\nu - \hat V^\mu \hat V^\nu )
+ (k_4+k_5) (\hat U^\mu \hat V^\nu + \hat U^\nu \hat V^\mu ) \Big )
(l _\mu - \hat T_\mu )_{;\nu}
    \nonumber \\
    &=&  -\frac{1}{4} \Big( ( k_3+k_6) P^{\mu \nu}
+ ( k_3-k_6) (\hat U^\mu \hat U^\nu - \hat V^\mu \hat V^\nu )
+ (k_4+k_5) (\hat U^\mu \hat V^\nu + \hat U^\nu \hat V^\mu ) \Big )
(l _\mu - \hat T^t K_\mu )_{;\nu}
    \nonumber \\
    &=&   - \frac{1}{4} ( k_3+k_6) \theta ^{(l)}
- \frac{1}{4} ( k_3-k_6) \sigma_+ ^{(l)} - \frac{1}{4} (k_4+k_5)
\sigma_\times ^{(l)} ,
\end{eqnarray}
where $\sigma_+ ^{(l)}, \sigma_\times ^{(l)}$ are the shears of $l^\mu$ defined by \cite{hawking1973}
\begin{equation}
     \sigma_+ ^{(l)} = (\hat U^\mu \hat U^\nu - \hat V^\mu \hat V^\nu ) l _{\mu ;\nu} \;\; \;\;\; \sigma_\times ^{(l)} = (\hat U^\mu \hat V^\nu + \hat U^\nu \hat V^\mu ) l _{\mu ;\nu} \;\;.
     \label{sisi}
\end{equation}

Therefore,
\begin{equation}
    \frac{1}{2} { h_\mu}^{\nu ;\mu} \hat N_\nu = - (\delta \hat N^\mu)_{;\mu}
- \frac{1}{2} k_1 \hat T^t \kappa - \frac{1}{4} ( k_3+k_6) \theta ^{(l)} - \frac{1}{4} ( k_3-k_6) \sigma_+ ^{(l)} - \frac{1}{4} (k_4+k_5)
\sigma_\times ^{(l)}.
\label{40}
\end{equation}

Finally, substituting Eq.(\ref{32}) and Eq.(\ref{40}) into
Eq.(\ref{differential 2}),
we find
\begin{eqnarray}
 \delta M = &&- \frac{1}{8 \pi} \int _{\partial\Sigma_\text{in} }
\biggl ( \delta \kappa + \frac{1}{\hat T^t} 
\Bigl(  \, \delta \theta^{(l)}+\frac{1}{4} ( k_3+k_6) \theta ^{(l)}
+ \frac{1}{4} ( k_3-k_6) \sigma_+ ^{(l)} + \frac{1}{4} (k_4+k_5)
\sigma_\times ^{(l)}  \Bigl) \biggl ) \; dA \nonumber \\   &&+ \frac{1}{8 \pi}
\int _{\partial\Sigma_\text{in} } \delta \kappa  \; dA
+ \frac{1}{8 \pi} \int _{\partial\Sigma_\text{in} }\!\!\!
\kappa  \; \delta (dA) \; .
\end{eqnarray}
Or in summary,
\begin{eqnarray}
 \boxed{\delta M = - \frac{1}{8 \pi} \int _{\partial\Sigma_\text{in} }
\biggl(  \, \delta \theta ^{(l)} + \frac{1}{4} ( k_3+k_6) \theta ^{(l)}
+ \frac{1}{4} ( k_3-k_6) \sigma_+ ^{(l)} + \frac{1}{4} (k_4+k_5)
\sigma_\times ^{(l)} \biggl) {\cal N} \; dA
+ \frac{1}{8 \pi} \int _{\partial\Sigma_\text{in} }
\!\!\!\kappa  \; \delta (dA).}
\label{firstlaw2Ap}
\end{eqnarray}
It is worth noting that 
\begin{eqnarray}
    \boxed{ \delta \theta ^{(l)}= -\frac{k_2}{2} \theta^{(l)} + \frac{1}{2} (k_3 + k_6)_{;\rho} \hat N^\rho }
    \label{theta0}
\end{eqnarray}
which separately depends only on $k_2, k_3, k_6$, and we will prove it next.

Since $P^{\mu \nu} \hat T_{\mu ; \nu} = P^{\mu \nu} (K_\mu \hat T^t)_{; \nu} = P^{\mu \nu} K_{\mu ; \nu} \hat T^t + P^{\mu \nu} K_\mu  (\hat T^t)_{; \nu} =0$, $ \theta ^{(l)}$ can be simplified as
\begin{eqnarray}
    \theta ^{(l)} = P^{\mu \nu} l_{\mu ; \nu}
= P^{\mu \nu} ( \hat T_{\mu ; \nu} + \hat N_{\mu ; \nu} )
= P^{\mu \nu} \hat N_{\mu ; \nu} .
\label{expan}
\end{eqnarray}
Thus the variation of $ \theta ^{(l)}$ is
\begin{eqnarray}
    \delta \theta ^{(l)} &=& \delta P^{\mu \nu} \hat N_{\mu ; \nu}
+  P^{\mu \nu} \delta (\hat N_{\mu ; \nu} ) 
    \nonumber \\
    &=& \delta P^{\mu \nu} \hat N_{\mu ; \nu} +  P^{\mu \nu} (\delta \hat N_{\mu , \nu} - \delta \Gamma^\lambda_{\mu \nu} \hat N_\lambda - \Gamma^\lambda_{\mu \nu} \delta \hat N_\lambda )
    \nonumber \\
    &=& \delta P^{\mu \nu} \hat N_{\mu ; \nu} +  P^{\mu \nu} \Bigl ( (\delta \hat N_\mu )_{; \nu} - \delta \Gamma^\lambda_{\mu \nu} \hat N_\lambda \Bigr )
    \nonumber \\
    &=& \delta P^{\mu \nu} \hat N_{\mu ; \nu} +  P^{\mu \nu} \Bigl ( (\frac{k_2}{2} \hat N_\mu )_{; \nu} -  \frac{1}{2} g^{\lambda \rho } ( h_{\mu \rho ;\nu } + h_{\nu \rho ;\mu }-h_{\mu \nu ;\rho } ) \hat N_\lambda \Bigr )
    \nonumber \\
    &=& \delta P^{\mu \nu} \hat N_{\mu ; \nu} +  P^{\mu \nu}(\frac{k_2}{2} \hat N_\mu )_{; \nu} - P^{\mu \nu} \frac{1}{2} ( h_{\mu \rho ;\nu } + h_{\nu \rho ;\mu }-h_{\mu \nu ;\rho } ) \hat N^\rho
    \nonumber \\
    &=& \delta P^{\mu \nu} \hat N_{\mu ; \nu} +  P^{\mu \nu}(\frac{k_2}{2} \hat N_\mu )_{; \nu} - P^{\mu \nu}  h_{\mu \rho ;\nu } \hat N^\rho + \frac{1}{2} P^{\mu \nu} h_{\mu \nu ;\rho } \hat N^\rho
    \nonumber \\
    &=& \delta P^{\mu \nu} \hat N_{\mu ; \nu} +  P^{\mu \nu}(\frac{k_2}{2} \hat N_\mu )_{; \nu} -  (\hat N^\rho h_{\mu \rho})_{ ;\nu } P^{\mu \nu} + \hat {N^\rho}_{; \nu} P^{\mu \nu} h_{\mu \rho} + \frac{1}{2} (P^{\mu \nu} h_{\mu \nu})_{;\rho } \hat N^\rho - \frac{1}{2} {P^{\mu \nu}}_{;\rho } \hat N^\rho h_{\mu \nu}
    \nonumber \\
    &=& \delta P^{\mu \nu} \hat N_{\mu ; \nu} +  P^{\mu \nu}(\frac{k_2}{2} \hat N_\mu )_{; \nu} -  (k_2 \hat N_\mu)_{ ;\nu } P^{\mu \nu} + \hat {N^\rho}_{; \nu} ( \delta (P^{\mu \nu} g_{\mu \rho} ) - \delta P^{\mu \nu} g_{\mu \rho} )+ \frac{1}{2} (k_3 + k_6)_{;\rho } \hat N^\rho
    \nonumber \\
    &=& \delta P^{\mu \nu} \hat N_{\mu ; \nu} -  P^{\mu \nu}(\frac{k_2}{2} \hat N_\mu )_{; \nu} - \delta P^{\mu \nu} \hat N_{\mu ; \nu} + \frac{1}{2} (k_3 + k_6)_{;\rho } \hat N^\rho
    \nonumber \\
    &=& - \frac{k_2}{2} P^{\mu \nu} \hat N_{\mu ; \nu} + \frac{1}{2} (k_3 + k_6)_{;\rho } \hat N^\rho 
    \nonumber \\
    &=& - \frac{k_2}{2} \theta ^{(l)} + \frac{1}{2} (k_3 + k_6)_{;\rho } \hat N^\rho ,
\end{eqnarray}
where we have used Eq.~(\ref{de00}) in the fourth line and ${P^{\mu \nu}}_{;\rho } \hat N^\rho h_{\mu \nu} = 0$ in the seventh line.

This completes the proof of Eq.~(\ref{theta0}).
\qed

\def\xyxy{
\vskip 0.3in

Next, we prove that the two shears $\sigma_+ ^{(l)}$ and
$\sigma_\times ^{(l)}$ may be rotated into each other 
and that a principle shear exists. Consider a rotation of the dyad
$\{\hat U^\mu,\hat V^\mu\}$ into $\{{\hat U}'^\mu,{\hat V}'^\mu\}$ via
\begin{eqnarray}
    {\hat U}'^\mu = \hat U^\mu \cos{\omega} - \hat V^\mu \sin{\omega},
\qquad
    {\hat V}'^\mu = \hat V^\mu \cos{\omega} + \hat U^\mu \sin{\omega} \; .
\end{eqnarray}
According to Eq.~(\ref{sisi}), the rotated shears become
\begin{eqnarray}
{\sigma'}_+ ^{(l)} &=& l _{\mu ;\nu} (\hat U^\mu \hat U^\nu
 - \hat V^\mu \hat V^\nu ) \cos{2 \omega} -
 l _{\mu ;\nu} (\hat U^\mu \hat V^\nu +
\hat U^\nu \hat V^\mu ) \sin{2 \omega} = \sigma_+ ^{(l)} \cos{2 \omega}
-  \sigma_\times ^{(l)} \sin{2 \omega}      \nonumber \\
{\sigma'}_\times ^{(l)} &=&  l _{\mu ;\nu}
(\hat U^\mu \hat V^\nu + \hat U^\nu \hat V^\mu ) \cos{2 \omega}
+  l _{\mu ;\nu} (\hat U^\mu \hat U^\nu
- \hat V^\mu \hat V^\nu ) \sin{2 \omega}
= \sigma_\times ^{(l)} \cos{2 \omega} +  \sigma_+ ^{(l)} \sin{2 \omega} \;.
\label{sisi2}
\end{eqnarray}
We may simplify these by choosing an angle $\Delta$ to give
\begin{eqnarray}
    {\sigma'}_+ ^{(l)} = {\sigma}_P^{(l)}
\cos{(2 \omega + \Delta) },\qquad
    {\sigma}_\times ^{(l)} = \sigma_P^{(l)}
\sin{(2 \omega + \Delta) },
\qquad {\text{where}}~~ \sigma_P^{(l)} \equiv \sqrt{{\sigma_+ ^{(l)}}^2 
+ {\sigma_\times ^{(l)}}^2} \;.
\end{eqnarray}
Now choosing $2\omega=\pi/2-\Delta$ we see that 
${\sigma'}_+ ^{(l)}=0$ and ${\sigma'}_\times ^{(l)}=\sigma_P^{(l)}$
which we dub the `principle shear'.
Note that the expansion $\theta ^{(l)}$ is invariant under this dyad
rotation. Therefore for a suitable orientation of the dyad 
$\{\hat U^\mu,\hat V^\mu\}$ at each point on the surface
$\partial \Sigma_{\text{in}}$, Eq.~(\ref{firstlaw2Ap}) may be reduced to
\begin{eqnarray}
 \delta M &=& - \frac{1}{32 \pi} \int _{\partial\Sigma_\text{in} }
\biggl(  4 \, \delta \theta ^{(l)} + ( k_3+k_6) \theta ^{(l)} + (k_4+k_5)
\sigma_\text{P}^{(l)} \biggl) {\cal N} \; dA
+ \frac{1}{8 \pi} \int _{\partial\Sigma_\text{in} }
\!\!\!\kappa  \; \delta (dA).
\label{firstlaw3}
\end{eqnarray}
On a horizon, both the expansion $\theta ^{(l)}$ and (principle) shear
$\sigma_\text{P}^{(l)}$ vanish \cite{hawking1973}, thus
Eq.~(\ref{firstlaw2Ap}) exactly reduces to  
\begin{equation}
    \delta M =  \frac{1}{8 \pi} \int _{\partial\Sigma_\text{BH} }
\!\!\!\kappa  \; \delta (dA).
    \label{fiiii}
\end{equation}

Further, as the expansion and shear vary linearly as one moves away
from the horizon, for a stretched horizon that is sufficiently close
to the horizon, the correction terms to the first law, Eq.~(\ref{fiiii}),
may be made negligibly small.

Holographic screen case

.............

.............

.............

.............

.............

.............
}

\def\xxy{
Although more general surfaces are ruled out since they would not
be in thermodynamic equilibrium with a physical screen, here we provide
a separate heuristic argument based on parameter counting.
Outside of a horizon, we have $\theta ^{(l)}>0$, so we may write
Eq.~(\ref{firstlaw3}) as
\begin{eqnarray}
 \delta M &=& - \frac{1}{32 \pi} \int _{\partial\Sigma_\text{in} }
\Bigl(  ( 4\,\delta \ln \theta^{(l)} 
+ k_3+k_6 ) \theta ^{(l)} + (k_4+k_5)
\sigma_\text{P}^{(l)} \Bigl)\, {\cal N} \, dA
+\frac{1}{8 \pi} \int _{\partial\Sigma_\text{in} }
\!\!\!\kappa  \; \delta (dA).
\label{firstlaw4}
\end{eqnarray}
As $\theta ^{(l)}$ and $\sigma_\text{P}^{(l)}$ are independent of each
other for arbitrary surfaces, in order to make the (unwanted) first
integral of Eq.~(\ref{firstlaw4}) vanish, we require that both
$4\,\delta \ln \theta^{(l)} + k_3+k_6$ and $k_4+k_5$ identically
vanish. However, the perturbed inner boundary can locally only impose
a single constraint on $h_{\mu\nu}$, thus in general the first law cannot
hold. One caveat to this is if the initial inner boundary is spherically
symmetric, in which case the shear is zero and only a single constraint
is required in order to satisfy the first law. Indeed, in the more
restrictive case of spherical symmetry in both the initial and 
final spacetime and initial and final inner boundary, this has already
been noted in Ref.~\onlinecite{Chen11}.\\
}

\end{widetext}
\subsection{Vanishing extrinsic curvature tensor}

Our analysis has been predicated on static screens. However, there is
another way to define screens, so their normal direction remains
parallel to the proper acceleration of a family of locally coincident
timelike observers \cite{Piazza2010}. These observers are constrained to
have constant 4-acceleration along with a number of other technical
assumptions \cite{Piazza2010}. A first law is then obtained for these
surfaces provided they additionally have a vanishing extrinsic curvature
tensor $K_{\mu\nu}=0$ \cite{Piazza2010}. The first law obtained is of a
form with energy and temperature measured locally instead of at spatial
infinity, which for asymptotically-flat spacetimes are unambiguous.
Finally, we note that there is no easy way in this other formalism 
\cite{Piazza2010} to investigate stretched horizons. 

In our setting with zero shift vector $\beta^\mu=0$, so $\hat T^\mu =
\hat T^t K^\mu$, and our hypersurfaces $\Sigma$ are orthogonal to $\hat
T^\mu$, we find that $K_{\mu\nu}=0$ implies a vanishing expansion
$\theta^{(l)}=0$. Thus, for our setting, the formalism of
Ref.~\onlinecite{Piazza2010} only yields a first law on horizons.

To see that this is the case, recall that the extrinsic curvature
tensor of our inner boundary equals \cite{carroll2004}
\begin{equation}
    K_{\mu \nu} \equiv \hat N_{(\lambda ; \rho)}
{P^\lambda}_\mu {P^\rho}_\nu.
\end{equation}
Taking the trace of this yields the extrinsic curvature scalar as
$K=P^{\mu\nu} \hat N_{\mu;\nu} =\theta^{(l)}$, where in the final step
we use Eq.~(\ref{expan}). Thus, for our setting, the first law of
Ref.~\onlinecite{Piazza2010} appears to occur at the horizon; a result
which is naively consistent with the classic 1973 result.

Let us now consider a construction for a screen surrounding a
gravitating body as proposed by Ref.~\onlinecite{Piazza2010}: Construct
a screen using a family of stationary timelike observers at fixed radius
around a Schwarzschild black hole. It is easy to calculate the extrinsic
curvature tensor for the screen and see, as noted above, that this
curvature vanishes only on the horizon. Hence the screen is on the
horizon and the observers are null instead of timelike observers. Next
drop in a spherical shell of matter. As the shell passes the screen of
observers, the horizon (where $\theta^{(l)}=0$) discontinuously jumps,
the surface gravity of the new horizon changes and the original screen
of observers fall into the black hole. We must then conclude either that
the construction using the methods of Ref.~\onlinecite{Piazza2010} of a
screen surrounding the black hole is simply impossible (because the
observers are not timelike), or it fails to continue to hold under
perturbation.

Thus, although Ref.~\onlinecite{Piazza2010} purports to describe a
dynamical first law for ordinary surfaces its conditions are either in
general impossible to satisfy or are generally {\it not\/} preserved
under perturbation.

\subsection{Local temperature in emergent gravity}

\begin{figure}[ht]
\centering
\includegraphics[width=0.39\textwidth]{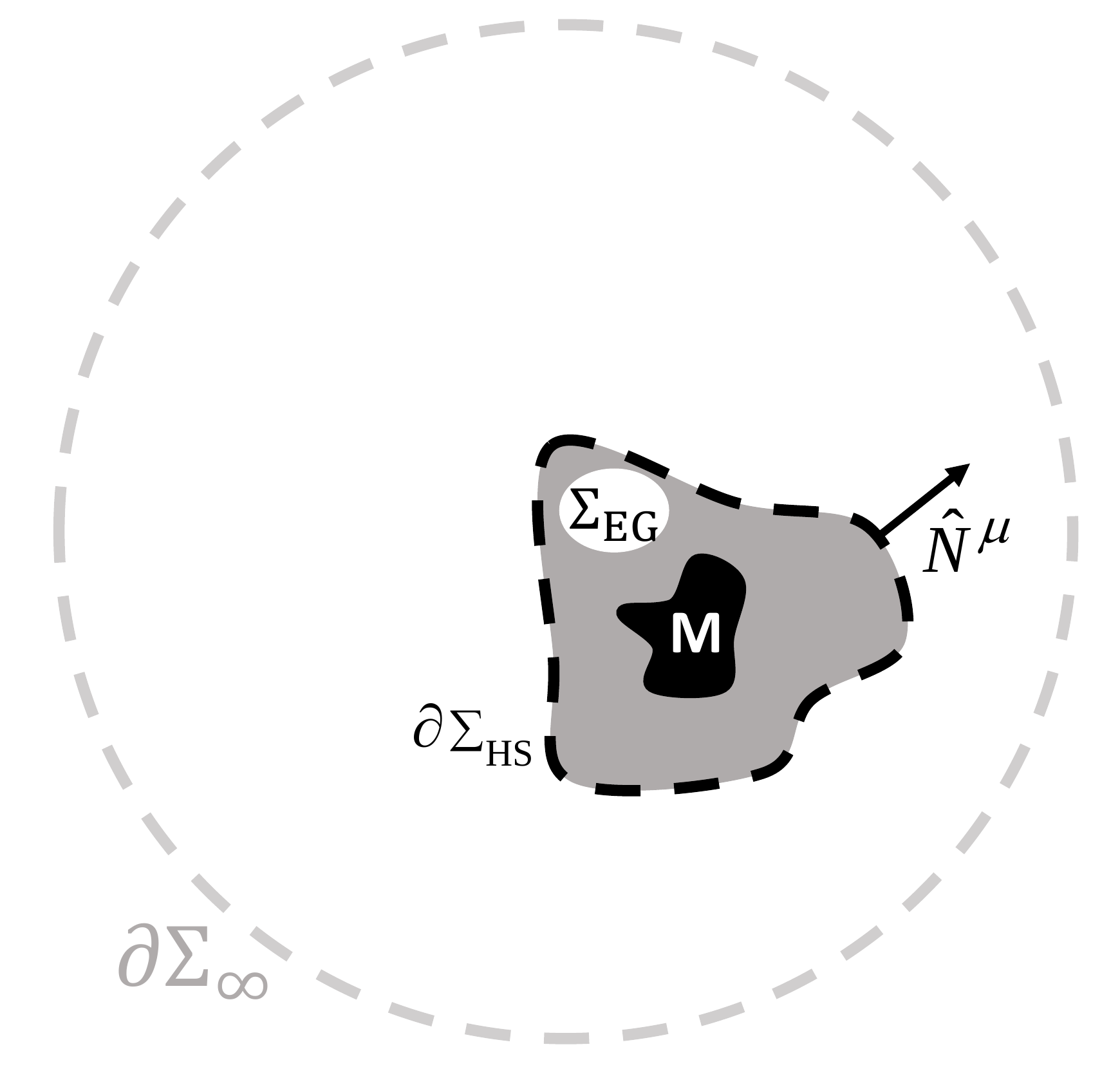}
\vskip -0.1truein
\caption{Schematic of the spacelike three-dimensional hypersurface
$\Sigma_{\text{EG}}$ used in Ref.~\onlinecite{Verlinde2011} which has
the mass under study embedded within it. As can be seen, the 2-surface
corresponding to the holographic screen $\partial \Sigma_{\text{HS}}$ is
now the {\it outer\/} boundary to $\Sigma_{\text{EG}}$
(compare to Fig.~\ref{fig1}); and Ref.~\onlinecite{Verlinde2011} defines
$\partial \Sigma_{\text{HS}}$ as ordinary surfaces of constant
Newtonian potential $\phi$.
(For context, we show spatial infinity as $\partial \Sigma_\infty$ in
grey, though it plays no roll in this section.)}
\label{fig2}
\end{figure}

We focus here on the temperature defined in the original paper on
emergent gravity \cite{Verlinde2011} which is used there in a heuristic
derivation of the Einstein field equations. In Fig.~\ref{fig2} we show a
schematic of the hypersurface considered there. $\partial
\Sigma_{\text{HS}}$ denotes the holographic screen (ordinary
surfaces of constant
Newtonian potential $\phi$) which now is the
outer boundary of the spacelike hypersurface $\Sigma_{\text{EG}}$ under
study, and $\hat N^\mu$ is the unit normal vector of the holographic screen.

\vskip -0.1truein

The `local' temperature of the holographic screen (as measured at
spatial infinity) used in Ref.~\onlinecite{Verlinde2011} is defined as
\begin{equation}
T\equiv \frac{1}{2\pi}e^\phi \,\phi_{;\mu} \hat N^\mu,
\end{equation}
where $\phi$ is the generalized Newtonian potential, given by
$\phi=\frac{1}{2} \text{ln} (-K^\mu K_\mu)=\ln {\cal N}$, recalling
that $K^\mu K_\mu=-{\cal N}^2$. It is now an easy matter to check that
\begin{eqnarray}
    T &\equiv& \frac{1}{2\pi}e^\phi \,\phi_{;\mu} \hat N^\mu 
=\frac{1}{2\pi}  {\cal N}_{;\mu} \hat N^\mu
= \frac{1}{2\pi} \frac{1}{2 {\cal N}}
({\cal N}^2)_{;\mu} \hat N^\mu \nonumber \\
    &=& -\frac{1}{2\pi} {K_\nu}_{;\mu}\frac{1}{\cal N}\,
K^\nu \hat N^\mu
= \frac{1}{2\pi} K_{\mu;\nu}\frac{1}{\cal N}\,
K^\nu \hat N^\mu \nonumber \\
    &=& \frac{1}{2\pi} K_{\mu;\nu}
\hat T^\nu \hat N^\mu . 
\end{eqnarray}
In summary, recall the definition of $\kappa$ in Eq.~(\ref{def}), yielding
\begin{equation}
    \boxed{T \equiv \frac{\kappa}{2 \pi}.}
\end{equation}

For reference, the Unruh temperature associated with a stationary observer
is just the {\it magnitude} of the observer's proper acceleration $a^\mu$ over
$2\pi$. As their 4-velocity is given by $\hat T^\mu$ we easily find
\begin{equation}
a^\mu \equiv \hbox{${\hat T}^\mu$}_{;\nu} {\hat T}^\nu=\phi^{;\mu},
\end{equation}
since ${\hat T}^\mu = {\hat T}^t K^\mu =K^\mu /{\cal N}=e^{-\phi}
K^\mu$. Thus $a^\mu$ is perpendicular to surfaces of constant $\phi$.
When Verlinde's temperature is measured locally (instead of
referenced to spatial infinity) it is $T_{\text{local}}=\frac{1}{2\pi}
\phi^{;\mu} {\hat N}_\mu$. For this to equal the Unruh temperature at
the same point, the local unit normal ${\hat N}^\mu$ to the screen must be
aligned with the proper acceleration $a^\mu$ of our stationary observer
there. Therefore, it trivially follows that only for surfaces of
constant Newtonian potential $\phi$ would the holographic screens be in
thermal equilibrium with stationary physical surfaces of the same shape,
size and location. Hence,
\begin{eqnarray}
    \boxed{
    \text{Thermodynamic equilibrium} \;\;
\Rightarrow \;\; \hat N^\mu \parallel \phi^{; \mu} }
\end{eqnarray}

Finally, we show that for surfaces of constant $\phi$, we have
$\delta\phi=k_1/2$.
Indeed, since ${\hat T}^t=1/{\cal N}=e^{-\phi}$, we have
\begin{equation}
\boxed{\delta\phi=-\frac{1}{{\hat T}^t} \delta {\hat T}^t
=\frac{1}{2} k_1,}
\end{equation}
where in the last step we have used Eqs.~(\ref{eq35}) and~(\ref{eq36}).


\end{document}